\documentclass[amssymb,prb,twocolumn,showpacs,superscriptaddress]{revtex4-1}
\usepackage{graphicx}
\usepackage{epsfig}
\usepackage{longtable}
\usepackage{amsmath}
\usepackage{amssymb}
\usepackage{bm}

\begin{document}


\title{Homogenous and heterogeneous magnetism in (Zn,Co)O}



\author{M. Sawicki} \email{mikes@ifpan.edu.pl}
\affiliation{Institute of Physics, Polish Academy of Sciences, al.~Lotnik\'{o}w 32/46, PL-02-668 Warszawa, Poland}

\author{E. Guziewicz} \email{guzel@ifpan.edu.pl}
\affiliation{Institute of Physics, Polish Academy of Sciences, al.~Lotnik\'{o}w 32/46, PL-02-668 Warszawa, Poland}

\author{M. I. {\L}ukasiewicz}
\affiliation{Institute of Physics, Polish Academy of Sciences, al.~Lotnik\'{o}w 32/46, PL-02-668 Warszawa, Poland}

\author{O. Proselkov}
\affiliation{Institute of Physics, Polish Academy of Sciences, al.~Lotnik\'{o}w 32/46, PL-02-668 Warszawa, Poland}

\author{I.~A.~Kowalik}
\affiliation{Institute of Physics, Polish Academy of Sciences, al.~Lotnik\'{o}w 32/46, PL-02-668 Warszawa, Poland}

\author{W.~Lisowski}
\affiliation{Institute of Physical Chemistry, Polish Academy of Sciences, ul.~Kasprzaka 44/52, PL-01-224 Warszawa, Poland}

\author{P.~D{\l}u\.zewski}
\affiliation{Institute of Physics, Polish Academy of Sciences, al.~Lotnik\'{o}w 32/46, PL-02-668 Warszawa, Poland}

\author{A.~Wittlin}
\affiliation{Institute of Physics, Polish Academy of Sciences, al.~Lotnik\'{o}w 32/46, PL-02-668 Warszawa, Poland}

\author{M.~Jaworski}
\affiliation{Institute of Physics, Polish Academy of Sciences, al.~Lotnik\'{o}w 32/46, PL-02-668 Warszawa, Poland}

\author{A. Wolska}
\affiliation{Institute of Physics, Polish Academy of Sciences, al.~Lotnik\'{o}w 32/46, PL-02-668 Warszawa, Poland}

\author{W.~Paszkowicz}
\affiliation{Institute of Physics, Polish Academy of Sciences, al.~Lotnik\'{o}w 32/46, PL-02-668 Warszawa, Poland}

\author{R.~Jakie{\l}a}
\affiliation{Institute of Physics, Polish Academy of Sciences, al.~Lotnik\'{o}w 32/46, PL-02-668 Warszawa, Poland}

\author{B.~S.~Witkowski}
\affiliation{Institute of Physics, Polish Academy of Sciences, al.~Lotnik\'{o}w 32/46, PL-02-668 Warszawa, Poland}

\author{L.~Wachnicki}
\affiliation{Institute of Physics, Polish Academy of Sciences, al.~Lotnik\'{o}w 32/46, PL-02-668 Warszawa, Poland}

\author{M.~T.~Klepka}
\affiliation{Institute of Physics, Polish Academy of Sciences, al.~Lotnik\'{o}w 32/46, PL-02-668 Warszawa, Poland}

\author{F.~J. Luque}
\affiliation{Depto. de F\'{\i}sica de la Materia Condensada, Universidad Aut\'{o}noma de Madrid, E-28049, Madrid, Spain}

\author{D. Arvanitis}
\affiliation{Department of Physics and Astronomy, Uppsala University, P.O. Box 516, 751 20 Uppsala, Sweden}

\author{J.~W. Sobczak}
\affiliation{Institute of Physical Chemistry, Polish Academy of Sciences, ul.~Kasprzaka 44/52, PL-01-224 Warszawa, Poland}

\author{M. Krawczyk}
\affiliation{Institute of Physical Chemistry, Polish Academy of Sciences, ul.~Kasprzaka 44/52, PL-01-224 Warszawa, Poland}

\author{A. Jablonski}
\affiliation{Institute of Physical Chemistry, Polish Academy of Sciences, ul.~Kasprzaka 44/52, PL-01-224 Warszawa, Poland}

\author{W.~Stefanowicz}
\affiliation{Institute of Physics, Polish Academy of Sciences, al.~Lotnik\'{o}w 32/46, PL-02-668 Warszawa, Poland}

\author{D.~Sztenkiel}
\affiliation{Institute of Physics, Polish Academy of Sciences, al.~Lotnik\'{o}w 32/46, PL-02-668 Warszawa, Poland}

\author{M.~Godlewski}
\affiliation{Institute of Physics, Polish Academy of Sciences, al.~Lotnik\'{o}w 32/46, PL-02-668 Warszawa, Poland}

\author{T.~Dietl}
\affiliation{Institute of Physics, Polish Academy of Sciences, al.~Lotnik\'{o}w 32/46, PL-02-668 Warszawa, Poland}
\affiliation{Institute of Theoretical Physics, Faculty of Physics, University of Warsaw, PL-00-681 Warszawa, Poland}


\date{\today}

\begin{abstract}
A series of (ZnO)$_m$(CoO)$_n$ digital alloys ($m=2,8,$~$n=1)$ and superlattices ($m=80$,~$n=5,10$) grown by atomic layer deposition has been investigated by a range of experimental methods. The data provide evidences that the Co interdiffusion in the digital alloy structures is sufficiently efficient to produce truly random Zn$_{1-x}$Co$_x$O mixed crystals with $x$ up to 40\%. Conversely, in the superlattice structures the interdiffusion is not strong enough to homogenize the Co content along the growth direction results in the formation of (Zn,Co)O films with spatially modulated Co concentrations. All structures deposited at 160$^o$C show magnetic properties specific to dilute magnetic semiconductors with localized spins $S = 3/2$ coupled by strong but short range {\em antiferromagnetic} interactions that lead to low temperature {\em spin-glass freezing}.

It is demonstrated that {\em ferromagnetic-like features}, visible exclusively in layers grown at 200$^\circ$C and above, are associated with an interfacial mesh of {\em metallic Co granules} residing between the substrate and the (Zn,Co)O layer. This explains why the magnitude of ferromagnetic signal is virtually independent of the film thickness as well as elucidates the origin of magnetic anisotropy, as observed by us and others.
Furthermore, in films grown at 200$^\circ$C and above we observe a {\em superparamagnetic} contribution.
In this case a sizable nanoparticle magnetic moment originates from ferromagnetic metallic Co or Co rich nanoparticles dispersed in the bulk of the layer.

Our conclusions have been derived for layers in which the Co concentration, distribution, and aggregation have been determined by:  secondary-ion mass spectroscopy, electron probe micro-analysis, high-resolution transmission electron microscopy with capabilities allowing for chemical analysis; x-ray absorption near-edge structure; extended x-ray absorption fine-structure; x-ray photoemission spectroscopy, and x-ray circular magnetic dichroism.
Macroscopic properties of the layers characterized by the above techniques have been investigated by superconducting quantum interference device magnetometery and microwave dielectric losses allowing to confirm the important role of metallic inclusions.
\end{abstract}

\pacs{75.50 Pp, 68.55.Nq, 75.75.Cd, 81.15.Gh}
\keywords{ALD, Ferromagnetic, ZnO, Co}

\maketitle


\section{Introduction}
\label{sec:Intro}
Since the theoretical suggestion by {\em {ab initio}} computations that (Zn,Co)O can be intrinsically ferromagnetic,\cite{Sato:2000_JJAP} and the subsequent experimental observation of high-temperature ferromagnetism,\cite{Ueda:2001_APL} this compound has reached the status of a model system for a broad class of dilute magnetic oxides (DMOs) and dilute magnetic semiconductors (DMSs), in which a robust ferromagnetism is observed despite a minute concentration of magnetic impurities. However, despite the apparent agreement between experiment and theory for many of such materials systems, it was rather soon realized that the origin of the abundant high-temperature ferromagnetism is far from being understood.\cite{Dietl:2003_NM} Indeed, over the recent years it has become more and more obvious that the understanding of these ferromagnets requires the use of advanced nanocharacterization tools in order to asses how magnetic impurities are actually incorporated and distributed depending on the growth conditions and co-doping.\cite{Bonanni:2010_CSR}
On the theoretical front,\cite{Dietl:2010_NM} it has been argued that strong correlation, disorder, and errors in the band gap, to mention only few challenges, make questionable the direct applicability of standard {\em ab initio} methods to these compounds.\cite{Zunger:2010_P}

Our view\cite{Bonanni:2010_CSR,Dietl:2010_NM} (not necessarily shared by all groups) in the context of the present work can be formulated as follows.

First, in the case of a random TM distribution and in the absence of valence band holes, no ferromagnetism is expected above $\sim$10~K.\cite{Dietl:1997_PRB,Dietl:2010_NM,Bonanni:2011_PRB}
In fact, in many studies, including our own, of (Zn,Mn)O (Refs.~\onlinecite{Fukumura:1999_APL,Kolesnik:2004_JAP,Lawes:2005_PRB,Wojcik:2006_APL,Wojcik:2007_APL}) and of (Zn,Co)O (Refs.~\onlinecite{Kolesnik:2004_JAP,Lawes:2005_PRB,White:2008_ChM,Ney:2010_PRB}) only a paramagnetic response has been observed, affected actually by  {\em antiferromagnetic} coupling between neighboring spins. This coupling is in fact so strong that the extrapolated N\'eel temperature would be close to 1000~K in wz-CoO. Such a robust antiferromagnetism, contradicting some {\em ab initio} predictions,\cite{Hanafin:2010_PRB} has indeed been found in colloidal nanocrystals of wz-Zn$_{1-x}$Co$_x$O studied up to $x = 1$.\cite{White:2008_ChM}

Second, the abundant observations of high temperature {\em ferromagnetism} in DMSs and DMOs, if not originating from experimental artifacts,\cite{Sawicki:2011_SST} are brought about by a highly non-random distribution of transition metal (TM) ions, introduced to the sample either purposely or \emph{via} contamination.\cite{Bonanni:2010_CSR}
This non-random distribution is driven by a significant contribution of $d$ levels to bonding which results in the formation of TM-rich nanocrystals characterized by high ordering temperature.

In the case of (Zn,Co)O, three origins of heterogeneous high temperature ferromagnetism have been considered.  First,  transmission electron microscopy,\cite{Park:2004_APL,Ney:2010_NJP} x-ray diffraction,\cite{Venkatesan:2007_APL} and x-ray magnetic circular dichroism\cite{Rode:2008_APL} give an evidence for the presence of metal Co inclusions. Also, the presence of superparamagnetic behavior in Co-implanted ZnO has also been assigned to Co nanoparticles, whose size could be changed by annealing.\cite{Zhou:2008_PRB} In our recent preliminary work, by employing depth profiling x-ray photoelectron spectroscopy, we have found that metallic Co is located at the interface to the substrate in films showing ferromagnetic features.\cite{Godlewski:2011_PSSB} Second, it has been suggested that the actual structure of relevant nanoparticles might be more complex, for instance, they could consist of an intermetallic ferromagnetic CoZn compound.\cite{Kaspar:2008_PRB} Finally, it has been argued that uncompensated spins at the surface of {\em antiferromagnetic} wz-CoO nanocrystals could give rise to spontaneous magnetization,\cite{Dietl:2007_PRB} the effect already observed in the case of NiO nanoparticles.\cite{Proenca:2011_PCCP}

In this paper, we present results of magnetic studies and of microwave dielectric losses measurements--sensitive to metallic inclusions---carried out on (Zn,Co)O samples characterized by a number of complementary element-specific probes.
Our films have been grown by atomic layer deposition (ALD) according to a procedure described in Sec.~\ref{sec:Samples}.
The ALD is a self-limiting growth process, introduced in the 1970s by Suntola and Antson (see Refs.~\onlinecite{Suntola:1977_USP,George:2010_ChR}) which had initially only limited niche applications (such as, {\em e.~g.}, thin film electroluminescence devices) but recently has proven itself highly successful and of rapidly growing technological importance, particularly when the Intel Company started to use ALD for deposition of a high-$\kappa$ oxide (HfO$_2$ and related oxides) as a gate dielectric in MOSFET transistors.
This method not only allows superior control of layer thickness and perfect coating of surfaces with different shapes and morphology, including the 3D ones, but, and most importantly for the present studies, allows a substantial reduction of the temperature of the deposition process.

When applied to (Zn,Co)O, ALD provides a high quality homogenous nanocrystalline structure with grains exhibiting a various degree of the orientation of the wurtzite (wz) $c$-axis.
Furthermore, this technique allows to grow superlattice-like structures (ZnO)$_m$(CoO)$_n$ with various average nominal values of Co concentrations $x_{\text{nom}} = n/(n + m)$ and periods $n$. As detailed in Secs.~\ref{sec:Samples}-\ref{sec:XMCD}, the Co concentration, distribution, and aggregation have been determined by:  secondary-ion mass spectroscopy (SIMS); electron probe micro-analysis (EPMA); x-ray absorption near-edge structure (XANES); extended x-ray absorption fine-structure (EXAFS); x-ray photoelectron spectroscopy (XPS); high-resolution transmission electron microscopy (HR-TEM) with capabilities allowing for chemical analysis, and x-ray circular magnetic dichroism (XMCD). In Sec.~\ref{sec:SQUID} we present results of magnetic measurements carried out employing a superconducting quantum interference device (SQUID) magnetometer. Section~\ref{sec:MW} contains results of microwave dielectric loss studies.

This set of experiments allow us to establish the relation between the growth parameters and the distribution of Co. Namely, when the growth process is carried out at 160$^\circ$C then independently of other growth details, layer thickness, and Co concentration the layers exhibit paramagnetic properties, that is in a sense that the other, 'legacy', DMSs did: (i) the level structure and magnetism of single TM ions can be described (including the $c$-axis related magnetic anisotropy) by the relevant group theoretical model; (ii) their paramagnetism is weakened by strong \emph{antiferromagnetic} superexchange among the nearest neighbor TM cations with no evidence for ferromagnetic coupling for any pair distances, and (iii) the samples freeze to a spin-glass state on lowering temperature. In particular, in the case of digital alloy superlattices (ZnO)$_m$(CoO)$_n$, $n =1$, the temperature and field dependencies of magnetization as a function of the Co content $x$ determined from the Curie constant show the behavior expected for randomly distributed spins coupled by short range antiferromagnetic interactions up to $x =40$\%. This indicates that interdiffusion homogenizes the Co distribution in these digital structures. In contrast, long period (ZnO)$_m$(CoO)$_n$ superlattices ($n = 5$ or 10) show properties which can be ascribed as ZnO/(Zn,Co)O superlattices. We conclude that interdiffusion is not strong enough to randomize the Co distribution along the growth direction in the long period superlattices.

Furthermore, we have found growth conditions leading to a ferromagnetic-like behavior.
More specifically, our results demonstrate that its presence is associated primarily with the growth temperature. When the growth temperature is risen to 200$^\circ$C or above -- the ferromagnetic-like response appears. It comes about in two main guises.

Firstly, as a robust, nearly temperature independent (specified by high $T_{\text{C}}$) and highly {\em ani}sotropic response which we unambiguously associate with a few nm thin metallic Co mesh located at the (Zn,Co)O/substrate interface. Its existence is widely documented throughout the paper and it explains why the magnitude of the ferromagnetic-like signal is virtually independent of the film thickness as well as elucidates the origin of magnetic anisotropy, as observed by us and others.\cite{Venkatesan:2004_PRL} Furthermore, it makes it possible to understand significant deviations from the standard superparamagnetic behavior visible in our samples as well as in many high-temperature DMSs and DMOs.\cite{Coey:2010_NJP}

Secondly, anomalous magnetic response is seen in the form of a superparamagnetic-like behavior, pointing to the presence of nanoparticles magnetized internally up to above room temperature.
We show by comparing the Co concentration from SIMS data with the one resulting from magnetic measurements that the relevant nanoparticles consist rather of ferromagnetic Co (or intermetallic compounds) than of uncompensated spins at the surface of aniferromagnetic CoO. Furthermore, at least so far, our studies do not provide hints for defect- or hydrogen-mediated ferromagnetic interactions.
Also, our data do not confirm the recently  suggested\cite{Straumal:2009_PRB} relation between ferromagnetism and density of grain boundaries in ZnO.

\section{Samples}
\label{sec:Samples}

\subsection{Growth method}
The (Zn,Co)O samples have been grown by ALD at substrate temperature between 160$^\circ$C and 300$^\circ$C employing the F-120 Microchemistry reactor and a double exchange mechanism of the growth. We use diethylzinc (DEZn) or dimethylzinc (DMZn) as a zinc precursor, cobalt(II) acetylacetonate (Co(acac)$_2$) as a cobalt precursor and deionized water as an oxygen precursor.
These highly reactive precursors are {\em sequentially} introduced to the growth chamber, so they meet only at the surface of the film. This very specific characteristic of the ALD technique means that our samples are deposited as (ZnO)$_m$(CoO)$_n$ periodic structures with various combinations of $m$ and $n$ values. We grow our layers in either a digital alloy fashion ($m=2$ or $8$; $n = 1$) or employing a superlattice concept ($m=80$ and $n=5$ or 10).
The studied films have been deposited on silicon, glass, and sapphire substrates, however as we find no qualitative differences among them we concentrate on Si-substrate based films.

Apart from above, a wide range of other parameters controls the ALD growth process - the length of the ALD pulses (the dwell time of the particular precursor in the growth chamber), the waiting time before the purging process by an inert gas, time of purging (by nitrogen here), growth temperature, etc.
This enables the preparation of films with  quite different crystallographic order, stoichiometry and Co content.

\subsection{Investigated samples}
\label{sec:samples}

\begin{table*}
\begin{tabular}{c c c r c c c c c c c c}
\hline
Sample & Growth & ZnO/CoO &\multicolumn{1}{c}d& $c$-axis & $x_{\text{Co}}$ & $x_{\text{Co}}$ &  $x_{\text{Co}}$ & $x_{\text{Co}}$&Electron & Comment \\
& T  & & & orientation & SIMS & EPMA & EDX\hspace{2mm} & SQUID &\hspace{0.3cm}concentration\hspace{0.3cm} & &\\
&[$^\circ$C] &  &[nm]& & [\%] &  [\%]&  [\%]&  [\%]&   [cm$^{-3}$]& \\
\hline
F53& 160& 8:1& 560& $\perp$ & 6.6 & 5.4 & 4.8 &5.0 (4.7)&$ 2.7\cdot 10^{17}$ & PM \\
F72& 160&80:10& 970 & $\parallel$ &	0.8& 0.74 &	0.8&0.7 (1.0)&$3.8\cdot 10^{18}$ & PM \\
F73& 160&80:5& 1020 & $\parallel$ &0.6& 0.62 & 1.2&0.7 (0.9)&$3.2\cdot 10^{18}$	& PM \\
F175& 160&8:1 & 760& random & 5.0 & 5.4 &	5.6&6.4 (6.6)&$1.4\cdot 10^{15}$ & PM \\
F176& 160&8:1 & 430& $\perp$ & 4.6 & -- &	5.0& 6.0 (5.9)& highly resistive & PM \\
F179& 160&2:1& 140 & random & 15(8.9)& -- &	10.6 &16 (14)&highly resistive & PM \\
F215& 160&8:1 & 680& $\perp$ & 4.1 & 4.9 &	4.6&4.4 (3.8)&$1.3\cdot 10^{17}$ & PM \\
F254& 160&8:1 & 70& random & 18-36& -- & 29 &42 (40)& highly resistive & PM \\
F268& 200&2:1 & 345 & random & 8 & -- & 9.0 & -- & $1.6\cdot 10^{16}$ &FM\\
F307& 200&2:1 & 250 & random & 8 & -- & 6.5 & -- & $2.2\cdot 10^{18}$ & FM \\
F309& 200&2:1 & 1210 & random & 7 & 4.9 & 5.4 & -- & $4.2\cdot 10^{18}$ & SP\\
F328& 200&2:1 & 60& random & 11.4 & -- & 39 & -- & $5.8\cdot 10^{18}$ & FM\\
F338& 200&2:1& 90& random & 2 & -- & 12 & -- & highly resistive & FM\\
\hline
\end{tabular}
\caption{List of the samples investigated in this study. We indicate the growth temperature, the ZnO to CoO cycles' ratio, thicknesses, wurtzite $c$-axis arrangement,  Co concentrations obtained from SIMS, EPMA, EDX, and SQUID (when applicable), and the Hall electron concentration. This is followed by the established layers magnetic character, when PM stand for purely paramagnetic layers (Sec.~\ref{sec:Paramagnetic}), SP denotes layers where in addition to PM a sizable superparamagnetic component dominates at elevated temperatures (Sec.~\ref{sec:BulkSP}), and FM indicates layers which additionally show a strong and temperature independent ferromagnetic response (Sec.~\ref{sec:FM-samples}).}
\label{tab:samples}
\end{table*}

In Table~\ref{tab:samples} we display pertinent parameters characterizing growth conditions and properties of samples investigated in this study.

\begin{figure}[b]
		\centering
        \includegraphics[width=8.4 cm]{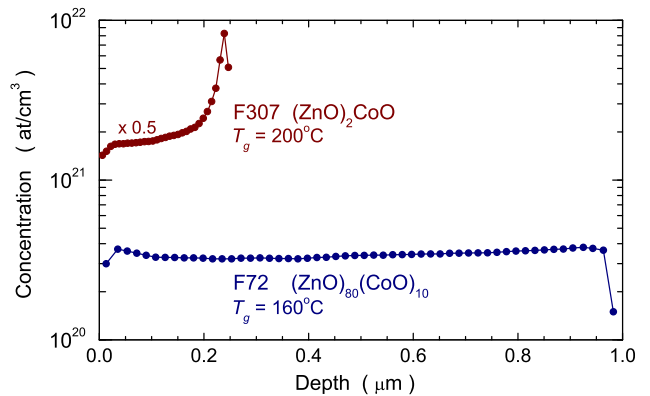}
      \caption{(Color online) SIMS Co depth profile for sample F72 and F307 deposited  at 160$^\circ$C and  200$^\circ$C, respectively. A Co accumulation at the interface to substrate is visible for sample F307.}
     \label{fig:SIMS}
\end{figure}

We find that it is the growth temperature of 200$^\circ$C and above that plays the decisive role in developing of ferromagnetic-like features in our layers.
Independently of the growth protocol, thickness of the layers, and an average Co content ferromagnetic and/or superparamagnetic  signatures are present there.
According to SIMS results presented in Fig.~\ref{fig:SIMS} for sample F307, a Co accumulation near the interface to substrate is visible in such samples.
This is corroborated by  TEM, XPS, XMCD data discussed in subsequent sections, which point to the presence of metal Co mesh at the interface.

The layers obtained at 160$^\circ$C do not show even slight signs of ferromagnetic or superparamagnetic features, they are as good paramagnet down to spin-glass freezing temperature as the Co-Co antiferromagnetic superexchange allows them to be.
A quantitative analysis presented in Sec.~\ref{sec:SQUID} demonstrates a perfectly random distribution of Co cations, with a clear distinction for $m=80$ and $n=5$ or 10 supperlattices, where insufficient out-diffusion into ZnO slabs  results in the ZnO/(Zn,Co)O superstructure instead of a uniform mixed alloy. This modulated Co distribution is beyond SIMS spacial resolution, which shows a uniform Co content along the growth axis for sample F72 (Fig.~\ref{fig:SIMS}).


Structural characteristics of the (Zn,Co)O films has been assessed by XRD measurements using the X-Pert MRD Pro Alpha1 diffractometer (Panalytical), equipped with an incident beam Ge monochromator and a strip detector.
Diffraction data are collected in a broad  2$\theta$ range (20$^\circ$ - 80$^\circ$), which includes main ZnO diffraction peaks corresponding to (10.0), (00.2), (10.1), and (11.0) crystallographic orientations.

\begin{figure}[b]
		\centering
        \includegraphics[width=8.5 cm]{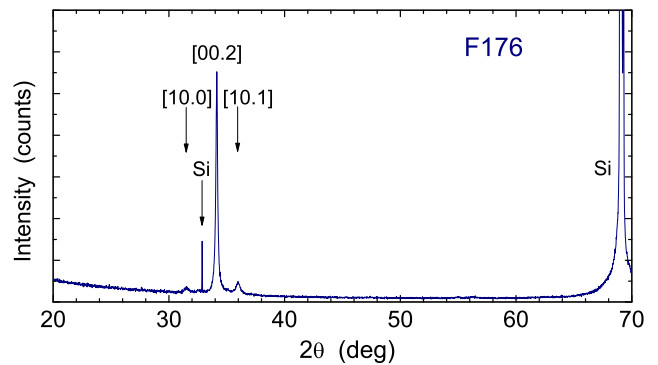}
      \caption{(Color online) XRD spectrum of layer F176. Indices of crystallographic directions corresponding to the diffraction maxima are indicated.}
     \label{fig:XRD_example_176}
\end{figure}

All the samples show a homogenous wurzite-type structure with a various degree of directional ordering of the wz $c$-axis.
We find no traces of foreign phases, cobalt oxides in particular, at least down to the base sensitivity of our equipment, estimated to be about ~1\%.
Such an exemplary XRD spectrum is shown in Fig.~\ref{fig:XRD_example_176}.
Here, a strong (00.2) reflex indicates that the majority of the (Zn,Co)O wz grains are oriented along the growth direction, whereas two weak (10.0) and (10.1) reflexes indicate that a certain, small,  fraction of the grains assume random orientations. From each spectra like this using Schrerrer's formula\cite{Schrerrer:1918_GN,Zsigmondy:1920_K}  we calculate an average grain size which stays within 5 to 7~nm for most of our layers. However, it assumes considerably larger values of nearly 30~nm in the case of F72 and F73 $m=80$ superlattices, where thick slabs of ZnO constitute a vast part of the layer.
This is in line with our previous findings for ALD grown (Zn,Co)O layers\cite{Kowalik:2009_JCG} and indicates a detrimental role of Co incorporation onto the crystallographic fidelity of (Zn,Co)O.

\begin{figure}[t]
		\centering
        \includegraphics[width=8.cm]{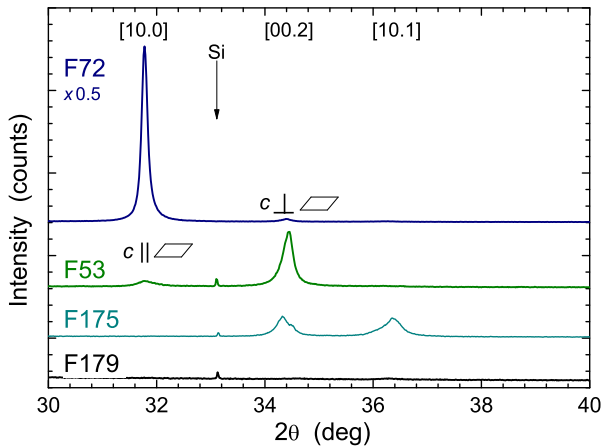}
      \caption{(Color online)  XRD spectra of four (Zn,Co)O layers showing ordered (F53 and F72) or random (F175 and F179) grains' orientation. The indices of crystallographic directions corresponding to the diffraction maxima as well as corresponding orientation of the wurtzite $c$-axis with respect to the sample plane are indicated.}
     \label{fig:XRD_Anisotropy}
\end{figure}

On the other hand, the ALD method allows to control the preferred orientation of the $c$-axis with respect to the substrate plane.\cite{Wojcik:2006_APL,Kowalik:2009_JCG}
As displayed in Table~\ref{tab:samples} and shown in Fig.~\ref{fig:XRD_Anisotropy} textured films with dominant orientation of the $c$-axis along the growth direction ({\em e.~g.},~F53 and F176), perpendicular to it ({\em e.~g.},~F72), or  polycrystalline films with  randomly oriented $c$-axis (samples F175 and F179) can be prepared.

There are however many conditions that influence the crystallographic orientation assumed by the grains. In addition to our previous studies of ZnO, where the purging time and growth temperature were identified as the decisive factors,\cite{Kowalik:2009_JCG} the present studies clearly show that an increase of the water pulsing time (the dwell time of the oxygen precursor in the ALD growth chamber) additionally favors the perpendicular orientation of grains whereas the presence of cobalt counters any grain order since most of the highly Co-concentrated samples show no preferred orientation. This is clearly a multidimensional parameter space and more dedicated studies are on the way to further refine this issue. However, as the $c$-axis orientation determines the type of the magnetic anisotropy exerted by the layers we come back to this issue in Sec.~\ref{sec:Mgn_Anisotropy} showing, in particular, that it is possible to extract a quantitative information on the $c$-axis distribution using low temperature magnetometry.

\section{XANES and EXAFS investigations}
\label{sec:XANES}

XANES and EXAFS experiments provide detailed and quantitative data on the local structure, therefore from this studies we extracted information about atomic configuration around Co ions and about a substitutional and/or interstitial position of cobalt in the ZnO matrix.
We perform comparable studies on a set of paramagnetic and ferromagnetic layers, represented below by F254 (paramagnetic) and F328 (ferromagnetic) (Zn,Co)O films.

\subsection{Methodology}
\label{sec:XstuffMethod}
XANES and EXAFS measurements at the K~edge of Co have been performed at DESY--Hasylab (Cemo station) at liquid nitrogen temperature using a 7-element silicon fluorescence detector.
Hard x-ray photons used in the experiment enabled penetration of the whole layer thickness.
The description of the experimental results is performed employing the IFEFFIT data analysis package making use of the Athena and Artemis codes.\cite{Ravel:2005_JSR} The passive electron reduction factor S02 = 0.8 is estimated from fitting the first shell of the data for the sample F254, where Co substitutes Zn.

Experimental results are compared to {\em ab-initio} computations carried out employing the FEFF 8.4 code\cite{Ankudinov:1998_PRB} in order to determine how Co atoms are located in the lattice. A cluster of 10~{\AA} radius is first created by using the known crystallographic data for possible Co-containing structures and then the XANES spectra are calculated employing the XANES, SCF (Self-Consistent Field) and FMS (Full Multiple Scattering) cards. The Hedin-Lundqvist exchange and correlation potential is adopted. Several structures are considered: cobalt oxides, ZnCo$_2$O$_4$, metallic cobalt, as well as (Zn,Co)O, where Co substitutes Zn.

\subsection{Experimental results - XANES and EXAFS}

\begin{figure}[t]
		\centering
        \includegraphics[width=8.5 cm]{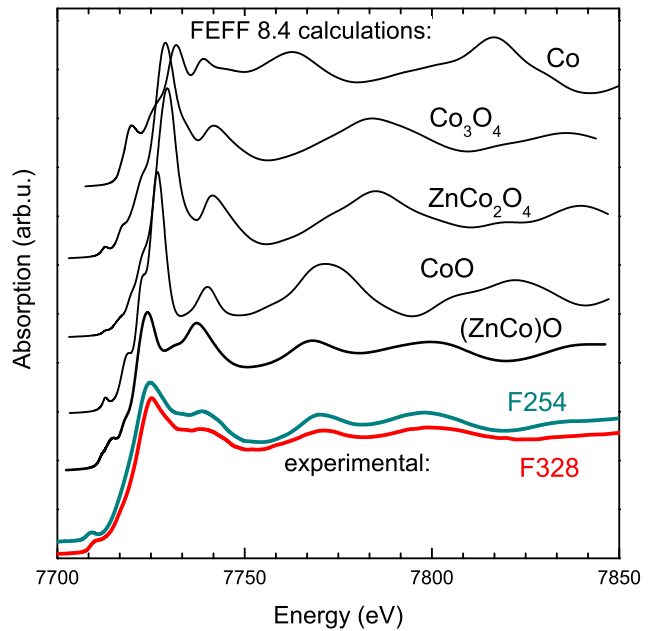}
      \caption{(Color online) XANES experimental spectra compared to results of computations for metallic Co and various compounds containing Co. Spectra are shifted vertically for clarity.}
     \label{fig:XANES}
\end{figure}

The experimental XANES spectrum compared to {\em ab initio} results is presented in Fig.~\ref{fig:XANES}. It can be noticed that for the sample F254 the measured spectrum shows the best agreement with the computed results for (Zn,Co)O. We conclude, therefore, that in the case of this sample Co atoms occupy cation substitutional positions. For the F328 sample the spectral features are slightly weaker indicating that a part of Co atoms can be in the form of other compound.

\begin{figure}[t]
		\centering
        \includegraphics[width=8 cm]{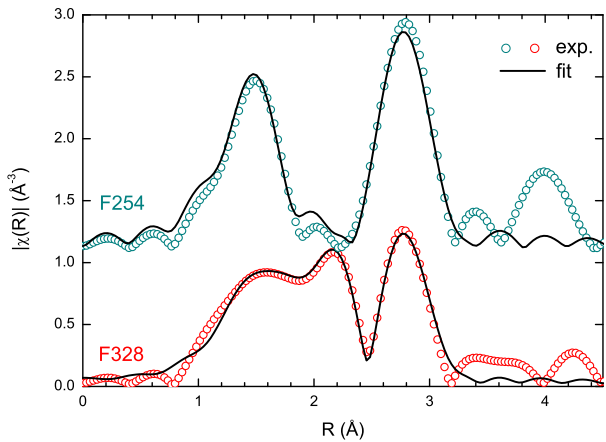}
      \caption{(Color online) Magnitude of the Fourier transform of the EXAFS spectra and fitting results for the samples. Spectra are shifted vertically for clarity.}
     \label{fig:EXAFS}
\end{figure}

In Fig.~\ref{fig:EXAFS}, the Fourier transforms of the EXAFS spectra and their fits are collected for the two studied samples. Fits parameters are displayed in Table~\ref{tab:EXAFS}.
As seen, the atom configurations around Co ion significantly differs in these two layers.
In the case of the sample F254, the first two maxima can be well described by coordination spheres expected for ZnO, {\em i.~e.},  four oxygen atoms at the distance of 1.97(2)~{\AA} and six zinc atoms at the distances of 3.19(1)~{\AA} and 3.23(1)~{\AA}, respectively.

However, in the F328 layer a new peak appears between the two maxima corresponding to the ZnO lattice. It is a clear indication for the presence of another crystallographic phase. In this case, in order to describe the experimental results, in addition to ZnO, other possible Co environments have been considered. It has been found that 73\% of (Zn,Co)O and 27\% of metallic Co gives the best description of the experimental data.

To conclude, XANES and EXAFS experiments show a different Co local structure for paramagnetic and ferromagnetic (Zn,Co)O samples. In paramagnetic films Co ions occupy substitutional Zn positions, whereas in ferromagnetic ones part of Co ions substitute zinc and a part of them are present in another crystallographic phase. Comparison with theoretical fitting indicates that the latter ones are in a metallic form.
To establish where metallic Co is located in the layer we have employed XPS and XPS profiling, discussed in the next section.

\begin{table}
	\begin{tabular}{c c c}
		\hline
		Parameter & F254 & F328 \\
		\hline
$R_{\text{Co-O}}$ [{\AA}] (4 atoms) & 1.97(1) & 1.98(2)\\
$ss^2_{\text{O}}$ & 0.005(2) &	0.007(3)\\
$R_{\text{Co-Zn1}}$ [{\AA}] (6 atoms) & 3.19(1)& 3.19(2)\\
$R_{\text{Co-Zn2}}$ [{\AA}] (6 atoms) & 3.23(1)&3.23(2)\\
$ss^2_{\text{Zn}}$ & 0.008(1) &	0.009(2)\\
$y$ [\%]& -- &	27(8)\\
$R_{\text{Co-Co}}$ [{\AA}] (12 atoms) & -- &2.49(2)\\
$ss^2_{\text{Co}}$ & -- &	0.005(3)\\
$R$-factor & 0.034 &	0.021\\
\hline
	\end{tabular}
	\caption{The fitting parameters to EXAFS data shown in Fig.~\ref{fig:EXAFS}. Here $R$ denotes the distance to the central atom; $ss^2$ is EXAFS Debye-Waller factor; $y$ is the relative contribution of metallic Co, and $R$-factor indicates the quality of the fit.}
\label{tab:EXAFS}
\end{table}

\section{XPS investigations}
\label{sec:XPS}

We have performed the XPS studies in order to determine the chemical state of cobalt in (Zn,Co)O films and to establish the location of Co within the layer. The XPS spectra are recorded on a PHI 5000 VersaProbe$^{TM}$ scanning ESCA Microprobe using monochromatic Al-K$_{\alpha}$ radiation ($h\nu  = 1486.6$~eV) from an x-ray source operating at 100~$\mu$m spot size, 25~W and 15~kV.
The high-resolution XPS spectra are collected with the analyzer pass energy of 23.5~eV, the energy step size 0.1~eV and the photoelectron take-off angle $45^\circ$ with respect to the surface plane. Shirley background subtraction and peak fitting with Gaussian-Lorentzian-shaped profiles are performed for the high-resolution XPS spectra analysis.
Binding energy (BE) scale is referenced to the C 1s peak with BE = 284.6~eV.

\begin{figure}[b]
		\centering
        \includegraphics[width=8.5 cm]{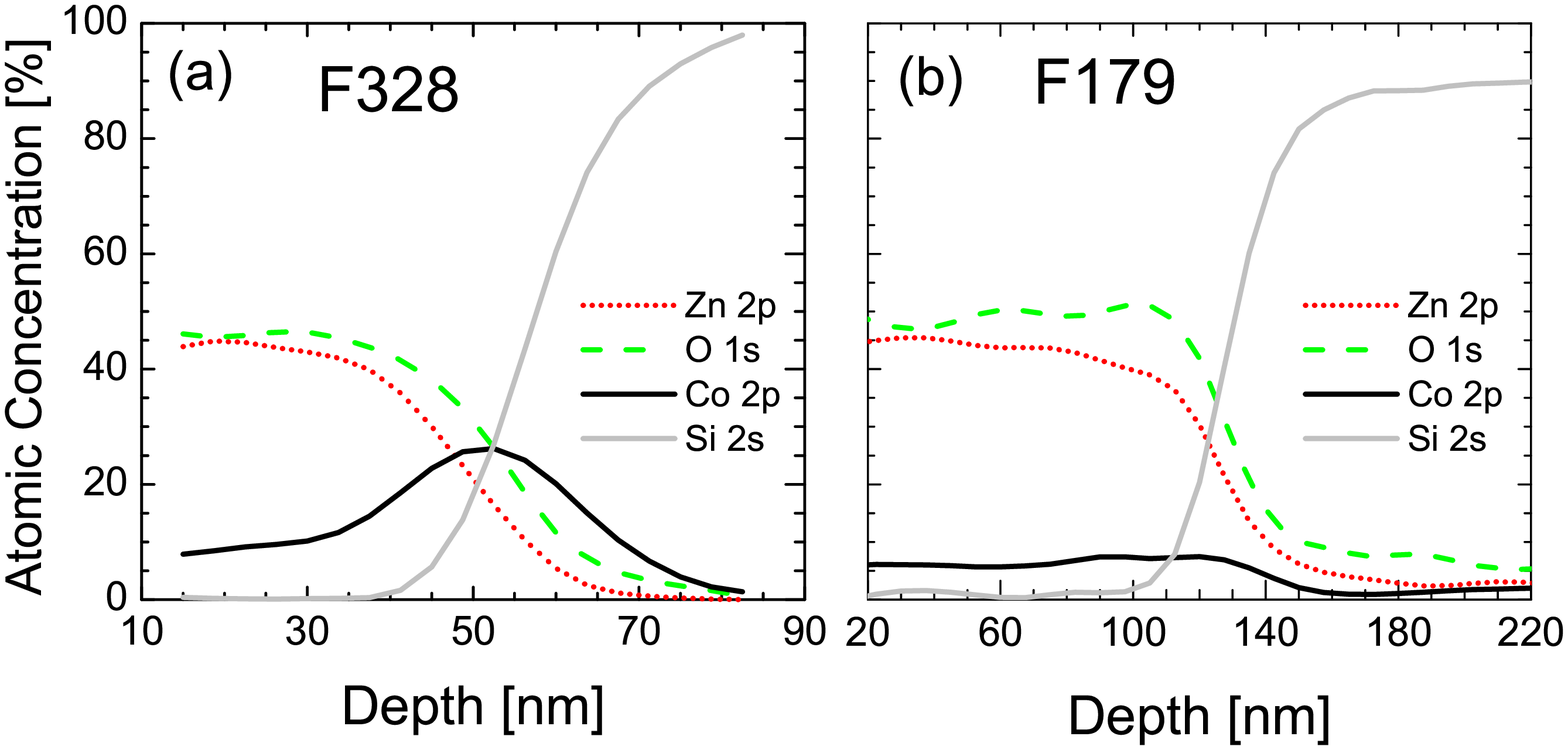}
      \caption{(Color online). XPS sputter depth-profiles of the a) ferromagnetic, b) paramagnetic (Zn,Co)O films.}
     \label{fig:XPS_1}
\end{figure}

\begin{figure}[t]
		\centering
        \includegraphics[width=10.5 cm]{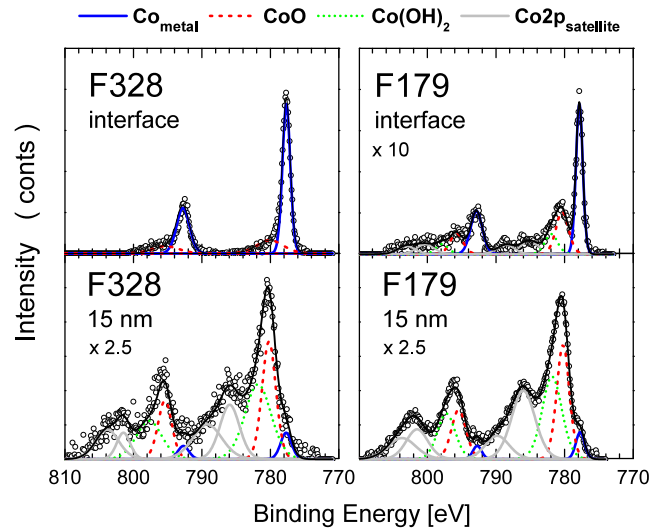}
      \caption{(Color online) Co2p core level XPS spectra measured during Ar$^{+}$ sputter profiling after removing of 15 nm of the (Zn,Co)O film (bottom) and at the (Zn,Co)O/Si interface region (top). Left: ferromagnetic (sample F328), right: paramagnetic (sample F179) (Zn,Co)O film. Deconvoluted spectra indicate different chemical states of Co compounds.}
     \label{fig:XPS_2}
\end{figure}

XPS is a surface sensitive technique in which the probing depth depends on the kinetic energy of photoelectrons, but in any case does not exceed few nanometers.\cite{Hufner:2003_B} In order to get information on the chemical state of cobalt within the bulk of investigated films we apply a depth-profiling XPS investigation in which XPS spectra have been recorded after sequential etching of the films by Ar$^{+}$ ions.
A few (Zn,Co)O films, both paramagnetic (PM) and ferromagnetic (FM), have been analyzed in this way.
In Fig.~\ref{fig:XPS_1} and Fig.~\ref{fig:XPS_2} we present the representative results for two (Zn,Co)O films: ferromagnetic (FM), (sample F328) and paramagnetic (PM), (sample F179).  The first 15~nm of the etched film was removed using 0.5~kV Ar ion etching with the rate of 1.5~nm per minute, and then the sputter rate of 15~nm per minute was used (2 kV Ar ion etching).
Both Ar$^{+}$  sputter rates have been measured using the SiO$_{2}$/Si reference sample.

As a result, we obtain elemental depth profiles that are presented in Fig.~\ref{fig:XPS_1}(a) and Fig.~\ref{fig:XPS_1}(b) for FM and PM films, respectively. All profiles reveal the elemental distribution in bulk of films from 15~nm depth until the (Zn,Co)O/Si interface was reached.
The relative atomic concentration of zinc, oxygen, cobalt, and silicon are evaluated from the intensity of XPS peaks associated with the Zn2p, O1s, Co2p and Si2s core levels.
Cobalt has been detected across all investigated (Zn,Co)O films.
However, its distribution is found to be different for PM and FM samples.
In the ferromagnetic (Zn,Co)O films a substantial enhancement of the Co concentration is found in the (Zn,Co)O/Si interface region, as shown in Fig.~\ref{fig:XPS_1}(a).
The cobalt concentration there is estimated to be more than three times larger than in the rest of the sample. Such a Co-rich interfacial layer is not observed in the paramagnetic (Zn,Co)O films [Fig.~\ref{fig:XPS_1}(b)].

In order to elucidate the chemical nature of cobalt compounds formed in the (Zn,Co)O films we analyzed the high-resolution XPS spectra of the Co2p, which have been taken at sequential steps of depth-profiling processing: at the surface of the (Zn,Co)O film, after removing of the 15~nm of the (Zn,Co)O layer from the top and at the (Zn,Co)O/Si interface. The last two results are presented in Fig.~\ref{fig:XPS_2}.

The original Co2p$_{3/2}$ spectra are presented together with deconvoluted peaks. The Co2p$_{3/2}$ and Co2p$_{1/2}$ spin-orbit contributions are separated by 15.0-15.2~eV.

The main Co2p contributions are accompanied by the broad shake-up satellites detected at larger BE.\cite{Naumkin:2008_NIST}
Following deconvolution analysis of the Co2p$_{3/2}$ XPS spectrum three forms of cobalt compounds can be distinguished in both samples at a depth of 15 nm (Fig.~\ref{fig:XPS_2}, bottom).

The strongest component, situated at BE = 780.3~eV, is characteristic for cobalt oxide.\cite{Naumkin:2008_NIST}
Two forms of cobalt oxides are stable in air at room temperature: CoO and Co$_3$O$_4$. They could be identified by the value of the Co2p spin-orbit splitting and by intensity of shake-up satellites and their positions relative to the main Co2p peaks \cite{Naumkin:2008_NIST,Chuang:1976_SS,Moulder:1995,Tan:1991_JACS}. The BE of the Co2p$_{3/2}$ peak reported for both cobalt oxides are very close.\cite{Naumkin:2008_NIST}
However, the XPS spectra of CoO are known to show strong satellite peaks above the main 2p line and the Co$_3$O$_4$ exhibits only weak satellites in this region.\cite {Tan:1991_JACS}
The Co2p spectra recorded during Ar$^{+}$  depth-profiling of both samples show strong satellites what indicate CoO to be a main cobalt oxide component. This means that the Co atoms that substitute Zn in the ZnO matrix, ZnO:Co, contribute to this component.

Another Co2p$_{3/2}$ XPS state is located at BE = 781.8~eV. Its chemical shift is characteristic of cobalt surrounded by -OH groups. The broad and high-intensive satellite peaks (BE at about 786~eV) are also indicative for a coexistence of -OH groups.\cite{Wagner:2003_NIST,Naumkin:2008_NIST,Tan:1991_JACS}

The Co2p$_{3/2}$ contribution at the BE=777.8~eV can be assigned to metallic cobalt.\cite{Naumkin:2008_NIST}
This observation indicates that some Co atoms in the (Zn,Co)O layers form metallic clusters and that these clusters are present both in ferromagnetic and paramagnetic films. Based on magnetic data it appears that in samples deposited at low temperatures the concentration and/or the size of Co inclusions is too small to be visible in magnetization measurements.

A significant difference between paramagnetic and ferromagnetic films is observed in the region of (Zn,Co)O/Si interface (see Fig.~\ref{fig:XPS_2}, top). In FM film we observe mainly metallic cobalt accompanied by very small concentration of CoO in the interface region. The absence of shake-up satellites, which is a characteristic feature of XPS spectra of metallic transition metals and rare earth, indicate the metallic Co to be a dominant chemical state and the cobalt oxide contribution is very small.  In the PM (Zn,Co)O film a metallic contribution is accompanied by cobalt oxides and Co-OH compounds, but also in this case the contribution from metallic cobalt is larger than in the volume of the sample.

During the Ar$^{+}$  depth-profiling experiment we have also recorded the O1s spectra (not shown here). Three, well-separated XPS states can be distinguished. The most intensive one at BE = 530~eV is attributed to oxygen bound to zinc. The peak at BE equal to 531.3~eV can be related to oxygen bound to Co. The lowest intensity peak, located at BE = 532.1~eV, can be assigned to the -OH groups. The apparent difference in the shape of the O1s peak of FM and PM films is observed in the interface region. The O1s XPS spectrum recorded from PM film   consists two nearly equaled contributions, one related to Co-O bonds, and second to -OH groups. For FM films oxygen bound to -OH groups dominates the XPS spectra. It is an open question what is the role of hydroxyl groups in the observed ferromagnetism of (Zn, Co)O.

In conclusion, the analysis of XPS spectra showed that inclusions of metallic cobalt clusters have been detected in both paramagnetic and ferromagnetic (Zn,Co)O samples. However, a different Co-cluster distribution has been observed within the (Zn,Co)O/Si interface region of both films. The interface in the ferromagnetic (Zn,Co)O film is formed mainly by accumulated metallic Co, whereas in the paramagnetic films the metallic Co is accompanied by cobalt oxides and hydroxides.



\section{HR-TEM investigations}
\label{sec:HR-TEM}

\begin{figure*}[t]
		\centering
        \includegraphics[width=14 cm]{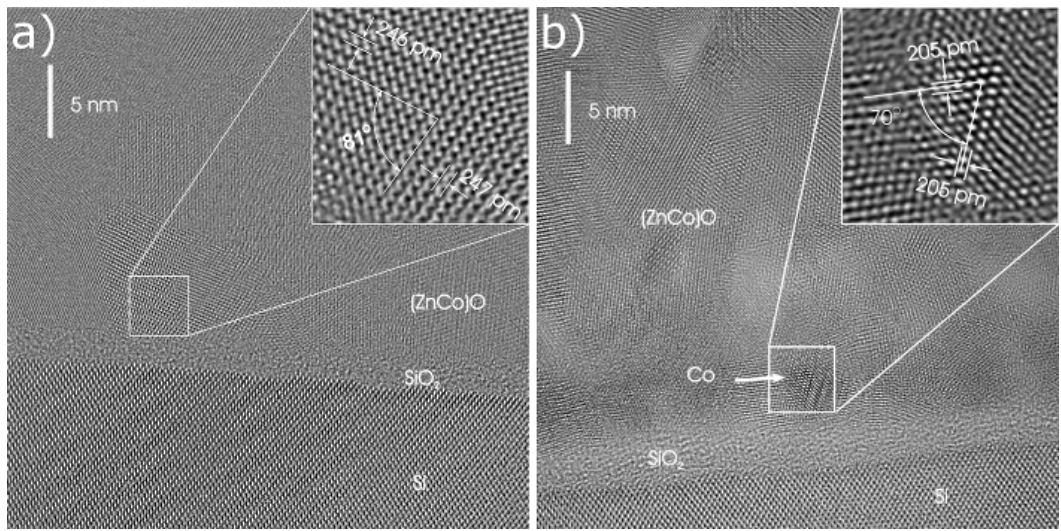}
      \caption{High resolution transmission electron microscopy  images of a) F179 sample with (Zn,Co)O wurtzite particle projected in the $\langle$01.1$\rangle$ direction,  and b) F328 sample with Co fcc nanoparticle projected in the $\langle$011$\rangle$ direction. }
     \label{fig:TEM}
\end{figure*}
Since both SIMS and XPS studies point to a presence of Co enriched regions of our layers we perform detailed high resolution transmission electron microscopy (HR-TEM) studies in cross section at 300~kV electron beam energy with the use of a Titan 80-300 Cubed Cs image corrected microscope equipped with an energy dispersion x-ray (EDX) spectrometer allowing for chemical analysis.
The cross sectional specimens have been prepared by ion milling proceeded with mechanical dimpling.
Only digital alloy (ZnO)$_m$/CoO layers have been investigated by TEM.
All the samples reveal a 2 - 3 nanometer thick amorphous SiO${_2}$ layer covering the Si(001) substrate, on which the (Zn,Co)O films are grown.

The HR-TEM microstructural analysis allows us to classify our layers into two main categories.
The first one, characteristic for layers grown at 160$^\circ$C, exemplified in Fig.~\ref{fig:TEM}a for layer F179, exhibits a uniform structure with oval, 5 to 10 nm in size, monocrystalline grains.
These values compare very well with the average grain size established from XRD.
All these grains show a wurzite structure characteristic of ZnO films as can be seen in the inset Fig.~\ref{fig:TEM}a.
This homogenous polycrystalline structure is observed at first 40 - 50 nm from the layer/Si interface.
In the next part of the film the columnar growth takes place and the width of the columns ranges from 25 nm to 30 nm.
No Co-rich volumes are found.
According to our extended magnetic studies (Sec.~\ref{sec:SQUID}) such layers exhibit paramagnetic properties typical for legacy DMS.

The second category, characteristic for layers grown above 160$^\circ$C, exhibits smaller oval crystallites (diameter of 3 - 4~nm) at the first 40 - 50 nm of the layer and narrower nanocolumns (10 - 22 nm in width) in the rest of the layer.
However, the basic difference between these two types of samples  lays in an existence of a layer of 3 to 4~nm in diameter Co clusters located at the layer/substrate, as  exemplified in Fig.~\ref{fig:TEM}b for layer F328.
The close-up on one of such clusters presented in the inset to Fig.~\ref{fig:TEM}b reveals the fcc crystallographic structure characteristic of metallic cobalt.
So, the HR-TEM characterization fully confirms conclusion derived from XPS studies about the large quantity of metallic cobalt present at the interfacial region of such films.
It is shown later (Sec.~\ref{sec:FM-samples}) that exactly this layer of Co clusters that coalesce to a form of metallic mesh spreading all over the layer/substrate interface is responsible for a robust, nearly temperature independent and highly anisotropic ferromagnetic response.

\begin{figure*}[t]
		\centering
        \includegraphics[width=14 cm]{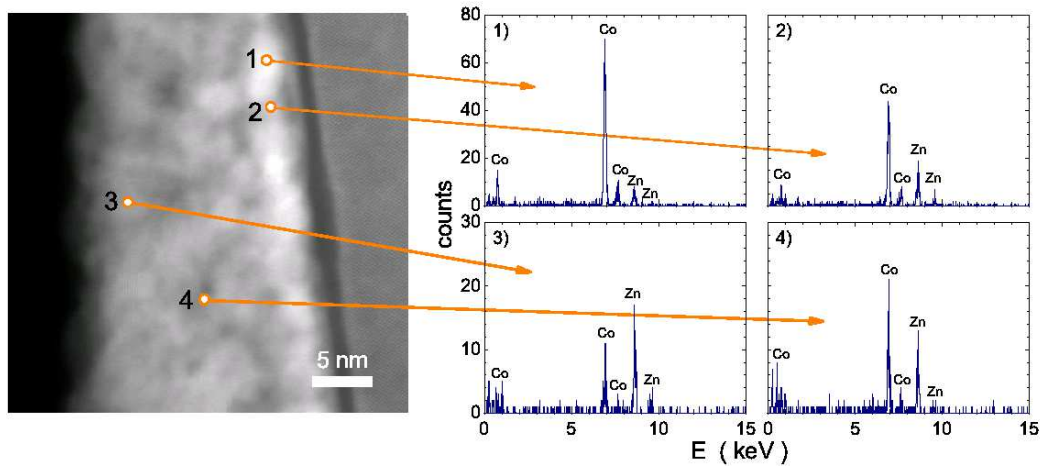}
      \caption{(Color online) A high angle annular dark field image of the sample F307 with indicated electron beam positions and corresponding to them EDX spectra.  The cobalt dominates on (Zn,Co)O/Si interface and exhibits high fluctuation inside the layer. }
     \label{fig:ChemAnalysis}
\end{figure*}
The point chemical analysis of other layer of this type presented in Fig.~\ref{fig:ChemAnalysis} gives evidence of the existence of Co rich volumes also in the bulk of the layer.
Samples like this one, as presented in Sec.~\ref{sec:BulkSP}, exhibit in an addition to the ferromagnetism stemming from the Co-rich interface region a clear superparamagnetic-like response postulated (see Sec.~\ref{sec:BulkSP}) to originate from about 2 - 3~nm in diameter hypothetical Co nanocrystallites.
We immediately note that such objects lay on the very border of the detection limit of our crossectional TEM method what probably accounts for a lack of their direct visualization  in our  HR-TEM images.
But, it is equally  possible that these small Co precipitations assume a coherent structure of the wurzite  surrounding.

\section{XAS and XMCD investigations}
\label{sec:XMCD}

\subsection{Experimental procedure}

We employ XMCD to verify in the chemical-sensitive fashion the origin of the observed ferromagnetic response in F328 sample at room temperature. The x-ray absorption spectroscopy (XAS) and x-ray magnetic circular dichroism (XMCD) measurements have been performed at I1011 beamline at MAX-lab synchrotron radiation facility in Lund, Sweden.\cite{Kowalik:2010_JPCS} This beamline is using an elliptically polarizing undulator source, producing soft x-rays of adjustable degree of polarization. It delivers high flux and high brightness circularly polarized x-rays in the energy range 0.2 to 1.7~keV, covering the L-edges of the late 3d elements. An ultra high vacuum end station has been employed in combination with this beamline, allowing for work in ultra high vacuum (UHV).
The samples have been mounted in the UHV chamber and characterized at room temperature by means of XAS and XMCD after baking the system for about 8 hours at 130$^\circ$C, to reach the base pressure of $5\cdot 10^{-10}$~mbar. A set of rotatable coils allows to apply a uniform magnetic field of 325~Oe, either in the direction perpendicular or almost parallel to the film plane, irrespectively of the x-ray angle of incidence. The magnetic coils are also employed to deliver magnetic field pulses of 650~Oe, between which the XMCD signal can be measured in the state of magnetic remanence. The measurements are performed in the total electron yield (TEY) mode by measuring the photocurrent of the sample. Using TEY, typically the near surface region, from the sample surface down to 6~nm below the sample surface is probed with the samples investigated here.	

\subsection{XAS and XMCD experimental results}

\begin{figure}[b]
		\centering
        \includegraphics[width=8 cm]{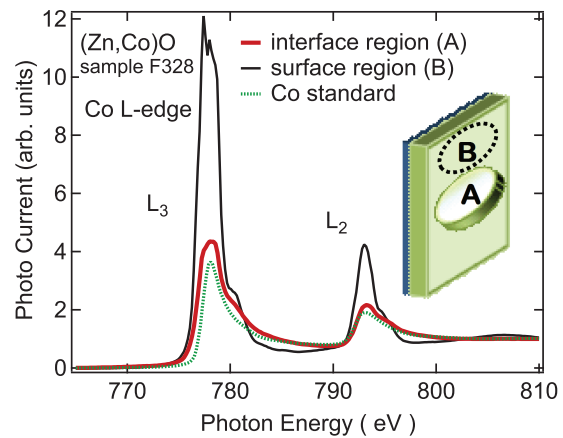} 
      \caption{(Color online)  L-edge x-ray absorption spectra versus photon energy for (Zn,Co)O from the interface region (region A, in the crater) and in the "as grown" state (region B of the as grown film); the Co L-edge data are taken at normal x-ray incidence (a 90$^\circ$ angle between the x-ray propagation direction and the film surface plane) in the total electron yield mode and normalized to the atomic continuum at high photon energies.}
     \label{fig:XMCD_1}
\end{figure}

First, by using the elemental specificity of XAS the composition of the (Zn,Co)O sample is probed for these sample positions. Here linearly polarized x-rays are used yielding a higher photon flux. The resonant character of the excitation makes XAS a more sensitive tool than non-resonant XPS or Auger-ES in the laboratory for checking the presence of impurities in small concentrations. The samples do not contain traces of Ti, V, Mn, Cr, Fe atoms. The L edges of these elements fall in the energy range we probe. The sample measured "as grown" in the first few nanometers contains about 6\% of Co ions on average. This amount can be calculated by using the tabulated atomic cross sections for XAS. The distribution of Co is inhomogeneous. The highest amount of Co measured on this sample is about 14\%, which is close to the value obtained from the analysis of the XANES data.
Also the fine structure of the Co white lines exhibits differences versus the "as grown" sample (Fig.~\ref{fig:XMCD_1}).

\begin{figure}[b]
		\centering
        \includegraphics[width=8 cm]{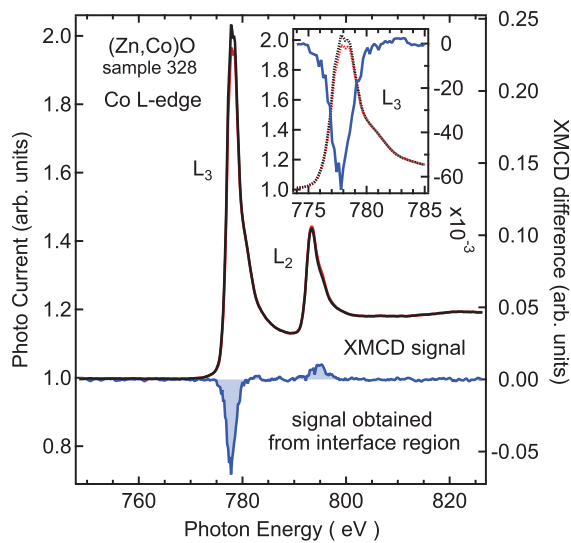} 
      \caption{(Color online) XAS (left scale) and XMCD (right scale) spectra in the total electron yield mode versus photon energy. Measurements are performed at room temperature under an applied magnetic field of $H = 350$~Oe. The XAS spectra are obtained with nearly fully circularly polarized light (a light helicity of 0.85); a 40$^\circ$  x-ray incidence angle is employed.  The insert shows an enlarged view of the L$_3$ edge of the XAS and XMCD spectra.}
     \label{fig:XMCD_2}
\end{figure}
The Co L-edge for the "as grown" film and around the crater shows the typical multiplet shape observed for (Zn,Co)O magnetically diluted samples.\cite{Kobayashi:2010_PRB}  Also the O K-edge is similar to the spectra shown in literature for (Zn,Co)O. However within the crater both the Co L-edge and O K-edges exhibit differences and indicate both a different stoichiometry as well as different electronic state for the O and Co ion cores probed by XAS. More in detail, by following the high energy atomic continuum it is seen that both the amount of Co and O does decrease within the crater. However the inter-peak continuum of final states for the Co spectra appears to increase within the crater. These final states, which exhibit \emph{s} symmetry, are clearly visible in the spectrum in the inter-peak region for metallic Co and can be used as a probe for the degree of metal character. The amount of the metallicity of the final states appears to increase within the crater, indicating that the Co may form there either a continuous ultrathin film or even small particles. The shape of XAS spectra obtained for Co L-edge indicates that in the interface region the metallic Co is the dominant phase but not the only one. We observe the superposition with small amount of CoO.\cite{Mulders:2009_JPCM} The difference of the electronic state of the Co atoms at the surface versus the interface region manifests itself also in the big difference in the number of holes of Co atoms in these two regions of the sample, illustrated as difference in the area under L$_3$ line (Fig.~\ref{fig:XMCD_1}). The number of holes for Co atoms in the surface region is about 7, in the crater is about 3.5, when for metallic Co this value is 2.8. This result confirms the difference in the hybridization of Co with neighboring atoms.  Inside the crater we not only observe a small amount of Co but also a small amount of Zn.

A small XMCD dichroic signal is seen in the interface region (inside the crater) at room temperature, when the magnetic field of 350~Oe is applied in the surface plane.  The dichroic response is much more pronounced at the L$_3$ white line indicating that the orbital moment carried by the Co atoms at the interface is much stronger than for bulk Co.
The magnitude of $m_l/m_s$ obtained by employing the magnetooptical "sum rules" for (Zn,Co)O is 0.31, whereas for metallic Co typical value is about 0.08. This is possible if the Co atoms agglomerate in the form of small particles with a metallic Co core.\cite{Tischer:1995_PRL,Gambardella:2002_N}
Alternatively, a very rough Co film in the crater with metallic patches is also consistent with our measured values.

Turning now to the overall value of the magnetic moment within the crater we note that at 300~K we obtain a value of the spin moment of 0.21 $\mu_{B}$ per Co atom, after correcting for the angle of x-ray incidence and the helicity of the x-rays. This value is of the order of 13\% of the spin moment for Co. We note in this context that a continuous film of only 2-3 atomic layers of Co should yield the full magnetic moment at 300~K, under the conditions of the present experiment. The low value of the magnetic moment which we obtain here is a further indication of stronger finite size effects than in the case of a thin film, indicating eventually Co particle formation.
We conclude that both the high orbital to spin moment ratio as well as the low value of the magnetic moment after the application of the magneto-optic sum rules indicate the presence of Co particles or a  discontinuous Co film within the crater area.

The value of the magnetic moments determined here are distinctively different from those found in ZnO containing Co particles of about 5~nm in diameter dispersed in the bulk of the films.\cite{Opel:2008_EPJB}
This difference illustrates the fact that the precise size, form and geometric shape of inclusions play a crucial role here and determines the outcome of an experiment. Conversely, an accumulated data base of such studies on well structurally characterized systems may later prove invaluable in the identification of the entities present in the investigated samples.

\section{SQUID investigations}
\label{sec:SQUID}
\subsection{Experimental procedure}
\label{sec:SQUID_ExpProc}

Having determined the Co distribution in our (Zn,Co)O layers we turn to investigations of macroscopic magnetic properties. In particular, we look for the nature and magnitude of interactions between dilute Co spins as well as for macroscopic manifestations of the Co aggregation observed at the nanoscale.
According to the data presented below, magnetization $M(T,H)$ of samples grown at 160$^\circ$C shows a paramagnetic (PM) behavior, which can be entirely described assuming that Co ions occupy only Zn-substitutional positions and that the spin-spin interaction is an antiferromagnetic short range superexchange, as in canonical II-VI Co-based DMSs. On the other hand, in samples grown at higher temperatures, in which nanocharacterization reveals metallic Co aggregation,  two additional contributions to $M(H)$ can also be observed: (i)  a fast saturating, highly anisotropic and temperature independent \emph{ferromagnetic} (FM) component and (ii) a relatively slowly saturating \emph{superparamagnetic} (SP) term. The manifestations of PM, FM, and SP contributions are discussed in subsequent sections.

All magnetic results presented here have been obtained with SQUID magnetometer (MPMS XL5 of Quantum Design) in its base temperature range: from 1.85~K (denoted here for simplicity as 2~K) up to about 390~K following the procedure described recently.\cite{Sawicki:2011_SST} In order to minimize contamination originating from tools and handling the specimens are rather cleft than cut. Routinely, we investigate about $5\times 5$~mm$^2$ specimens of a roughly square shape which are mounted either on a single 20~cm long and 1.5~mm wide silicon strip or between two such strips for in-plane and perpendicular to plane SQUID measurements, respectively.   The sample holding strips are customary cut from a commercial silicon wafer, provide adequate rigidity and stability, and make the measurements basically artifacts free at the whole accessible temperature $2< T <400$~K and  magnetic field $H$ ($\pm 50$~kOe) range. To save on the apparatus time and resources (helium), after careful checking for correctness, most of the field dependent measurements are performed during a one-way field sweeps only, and the missing return field sweep and so a full hysteresis loop (if needed for presentation) is created by numerical inversion of this one-way-sweep data set. When shown, this artificial data are marked differently.

\begin{figure}[b]
		\centering
        \includegraphics[width=8.4 cm]{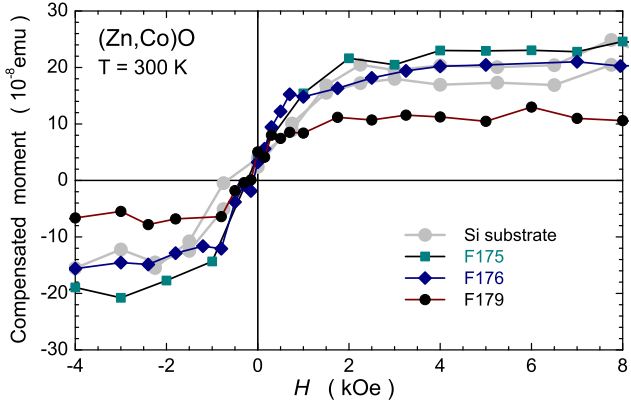}
      \caption{(Color online)  Nonlinear part of the magnetic response at  room temperature of paramagnetic (Zn,Co)O films on Si substrates. The magnetic field has been swept only once for each sample, from about +40 to -30 kOe. The combined diamagnetic signal from the substrate and paramagnetic of free Co cations has been derived for each sample by the slope of its $m(H)$ at large magnetic fields and subtracted from the original data. The compensation level for these measurements approaches 99.5\% already at $H=10$~kOe.}
     \label{fig:SQUID_1}
\end{figure}
All the films studied here have been deposited on a silicon substrate whose magnetic response constitutes a dominant part of the measured magnetic moment of the whole structure.
We have found that the substrate magnetization consists of the expected diamagnetic component linear in the magnetic field and of an unexpected but rather tiny nonlinear contribution, shown in Fig.~\ref{fig:SQUID_1}.
Whether this sigmoidal dependencies, suggesting some contamination by ferromagnetic nanoparticles, is intrinsic to the Si wafers employed or, less probably, results from a sample handling procedure, is unknown at present.
The shape of this signal is additionally perturbed by a trapped magnetic field in windings of the SQUID superconducting magnet.\cite{Sawicki:2011_SST}
A typical magnitude of this field after returning back to "zero" from 50~kOe, is 20~Oe. For a usual magnetic response of substrates, this field produces a spurious signal of the magnitude comparable to envisaged values of remanence in DMS films.

Magnetization values of (Zn,Co)O films presented in the subsequent subsections have been obtained by subtracting from the total signal the substrate contribution, measured as a function of the magnetic field and temperature, and adjusted for a particular sample by its  weight and the presence of the trapped field.

\subsection{Magnetization of Zn-substitutional Co ions}
\label{sec:Paramagnetic}

\subsubsection{Theoretical modeling}
\label{sec:SQUD_theory}
Following previous extensive studies of various DMSs and DMOs,\cite{Kolesnik:2004_JAP,Spalek:1986_PRB,Bonanni:2011_PRB} we exploit magnetic measurements in order to evaluate the  concentration distribution of Co ions as well as the character and magnitude of the exchange interactions between them.
In this section we summarized theoretical models and quote values of material parameters employed in subsequent sections to describe theoretically  the field and temperature dependence of magnetization $\bm{M}(T,\bm{H})$ brought about by Zn-substitutional Co ions.

Properties of a single Zn-substitutional Co$^{2+}$ ion ($d^7$) in wz-ZnO  can be described by the general $S=3/2$ Hamiltonian \cite{Estle:1961_BAPS}:
\begin{equation}
\label{eq:SpinHamiltonian}
 \mathcal{H} = g_{||}\mu_BH_zS_z+g_{\perp}\mu_B(H_xS_x+H_yS_y)+DS_z^2,
\end{equation}
where the $z$ axis of the coordinate system coincides with the hexagonal $c$-axis and ${\bm{H}}$ is an external magnetic field. The energy level positions are calculated by numerical diagonalization of the $4 \times 4$ Hamiltonian matrix $ \langle S_z|\mathcal{H}|S_z' \rangle $ with $|S_z \rangle=|-3/2\rangle,|-1/2\rangle,|1/2\rangle,|3/2\rangle$. Having eigenvalues, the partition function $Z$ and then magnetization  $\bm{M}(T,\bm{H})$ the concentration $xN_0$ of magnetic ions is obtained as,
\begin{equation}
\bm{M}(T,\bm{H}) =k_B T x N_0\partial \ln Z/\partial \bm{H},
\label{eq:M(T,H)}
\end{equation}
This general formula is equivalent to:
\begin{equation}
\label{eq:Magnetization}
\bm{M}_{x,y} = \mu_Bg_{\perp}  \langle \hat{S}_{x,y} \rangle, \bm{M}_{z} = \mu_Bg_{||}  \langle \hat{S}_z \rangle, \\
\end{equation}
where
\begin{equation}
\langle \hat{S} \rangle = \frac{\sum_{j=1}^4 \langle\varphi_{j}|\hat{S}|\varphi_{j} \rangle\mathrm{exp}(-E_j/k_BT)}{\sum_{j=1}^4\mathrm{exp}(-E_j/k_BT)}.
\end{equation}
Here $E_j$ and $\varphi_{j}$ are the $j$-th energy level and the eigenstate,  respectively.
This procedure allows us to calculate magnetic response of a Co$^{2+}$ ion,
which owing to a non-zero orbital momentum associated with the spin $S = 3/2$ of Co$^{2+}$ cations, depends on the angle $\theta$ between the magnetic field and the $c$ axis. Electron paramagnetic resonance (EPR),~\cite{Estle:1961_BAPS,Jedrecy:2004_PRB}, optical absorption,\cite{Koidl:1977_PRB} and direct magneto\-metry\cite{Sati:2006_PRL,Sati:2007_PRL, Ney:2010_PRB} prove the existence of a sizable low temperature magnetic anisotropy in Zn$_{1-x}$Co$_x$O.

We describe our data employing the values of parameters determined experimentally for Zn$_{1-x}$Co$_x$O: the cation concentration $N_0 = 4.2 \times 10^{22}$~cm$^{-3}$, the Land\'e factors $g_{\parallel} = 2.238$ ($H\parallel c$), $g_{\perp} = 2.276$ ($H\perp c$), and the uniaxial single-ion anisotropy energy $D = 0.342$ meV ($\simeq$ 4~K).\cite{Sati:2006_PRL,Ney:2010_PRB}
The positive sign of $D$ results in an 'easy-plane' configuration, that is (Zn,Co)O magnetizes faster with $H$ applied perpendicularly to the $c$-axis of the crystal.

In order to take into account in the theoretical modeling of $M(H)$ the presence of short range antiferromagnetic spin-spin interactions,  we follow the well established procedure.\cite{Gaj:1979_SSC,Dietl:2000_S}
It consists of replacing $x$ and $T$ in Eq.~\ref{eq:M(T,H)} by two temperature dependent fitting parameters, $x_{\text{eff}}$ and $T + T_{\text{AF}}$, respectively, where $T_{\text{AF}}> 0$ and $x_{\text{eff}} < x$. Typically, at helium temperatures, where the nearest neighbor spins form antiferromagnetically coupled magnetically inert singlet pairs,  $x_{\text{eff}} = x(1-x)^{z_1}$, where $z_1 = 12$ is the number of the nearest neighbor positions in the first coordination sphere in the cation wz-sublattice.\cite{Behringer:1958_JCP}

As noted in Sec.~\ref{sec:samples}, our films show a distribution of $c$ axis orientations. For polycrystalline samples, with a uniform distribution of $c$ axis, we average $M(H)$ over the full angle, corresponding to all orientations of the $c$ axis with respect to the magnetic field direction.

In order to describe textured films, in which there is a partial ordering of $c$ axis directions, we introduce an additional fitting parameter $y$ that describes the fraction of the crystal grains having their $c$-axis perpendicular to the sample plane. In terms of this parameter,  the magnitude of magnetization for the in-plane magnetic field assumes the form,
\begin{equation}
M(H \parallel \mbox{plane}) = y \cdot M(H \perp c) + (1-y)\cdot \langle M(H)\rangle_{\theta},
\label{eq:H_parallel}
\end{equation}
where $\langle M(H)\rangle_{\theta}$ denotes magnetization average over the in-plane angle ${\theta}$ between the direction of the magnetic field and uniformly distributed orientations of the $c$ axis in the film plane.

For the magnetic field perpendicular to the sample plane we employ
\begin{equation}
M(H \perp \mbox{plane}) = y \cdot M(H \parallel c) + (1-y) \cdot M(H \perp c).
\label{eq:H_perpen}
 \end{equation}

Next, we consider the temperature dependence of magnetic susceptibility. The presence of spin-spin interactions can be taken into account within the high temperature expansion.\cite{Spalek:1986_PRB} This procedure leads to the Curie-Weiss law which, for a random distribution of localized spins, is parameterized by two constants, $C_0$ and $\Theta_0$ independent of the Co concentration $x$,
\begin{equation}
 \chi(T) = xC_0/(T -x\Theta_0),
 \label{eq:CW}
 \end{equation}
where
\begin{equation}
 \Theta_0 = -\frac{1}{3}S(S+1)\sum_j z_jJ_j.
 \label{eq:Theta}
 \end{equation}
Here, the summation extends over the subsequent cation coordination spheres; $z_j$ is the number of cations in the sphere $j$, and $J_j \equiv J_{ij}$ is the corresponding Co-Co exchange integral in the Hamiltonian $H_{ij} = -J_{ij}\bm{S}_i\bm{S}_j$. We note that in another convention twice smaller values of $J_j$ are considered and then the factor  1/3 in Eq.~\ref{eq:Theta} is replaced by 2/3. Furthermore, an effective nearest neighbor exchange energy $J_{\text{eff}}$ is sometimes introduced in the literature, in terms of which
\begin{equation}
\Theta_0 = -\frac{1}{3}S(S+1)z_1J_{\text{eff}}.
\end{equation}

The values of the Land\'e factors quoted above point to an average value of the Land\'e factor $\langle g\rangle =2.25$, suitable to describe polycrystalline samples.
For this magnitude of $\langle g\rangle$  we obtain the Curie constant $C = xC_0$, where $C_0 =0.17$~emu\,K/cm$^3$.

The high temperature expansion is valid as long as $T \gg |\Theta|$, where the Curie-Weiss temperature $\Theta = x\Theta_0$. In a wider temperature range, $\chi(T)$ of random antiferromagnets is well described by\cite{Bhatt:1986_PS,Anderson:1986_PRB}  $\chi(T) = aT^{-\alpha}$,  where $a$ is a temperature independent constant and $\alpha < 1$.

As we show below, this time-honored model of magnetism in II-VI DMSs describes quite satisfactorily digitally doped samples grown at sufficiently low temperatures. It also applies to superlattice-like films provided that the modulated character of the Co distribution is taken into account.

\subsubsection{Temperature dependence of magnetic susceptibility}
\label{sec:ParaSamples}

\begin{figure}[b]
		\centering
        \includegraphics[width=8.4 cm]{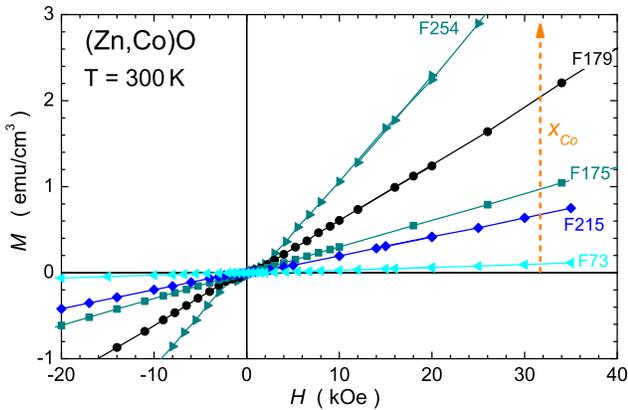}
      \caption{(Color online) An example of room temperature $M(H)$ for some of the paramagnetic (Zn,Co)O layers. The arrow indicates the direction of the growth of the Co content  in these layers according to the data accumulated in Table~\ref{tab:samples}. }
     \label{fig:SQUID_2}
\end{figure}
Figure~\ref{fig:SQUID_2} presents room temperature dependence of (volume) magnetization $M(H)$  on the applied field $H$ for selected (Zn,Co)O layers, where $M$ is obtained from $m$ using the whole thickness of the structure to assess the volume of the investigated layer.
All data show a good linearity in $H$, except the low field region where a residual sigmoidal signal, not completely removed by the subtraction procedure described above, mar the otherwise paramagnetic $M(H)$. Despite this little technical glitch, these data demonstrate that ferromagnetic or superparamagnetic contributions are negligibly small in these films. Samples exhibiting such properties are denoted as "PM" in Table~\ref{tab:samples}. We note that if a sample exhibit such a linear $M(H)$ at room temperature then \textsl{no} traces of ferromagnetic component are seen down to the base temperature of the magnetometer, $T \approx 2$~K.
In other words, if a ferromagnetic response is present, it persists to well above room temperature independently of a nominal Co concentration.


\begin{figure}[t]
		\centering
        \includegraphics[width=8.4 cm]{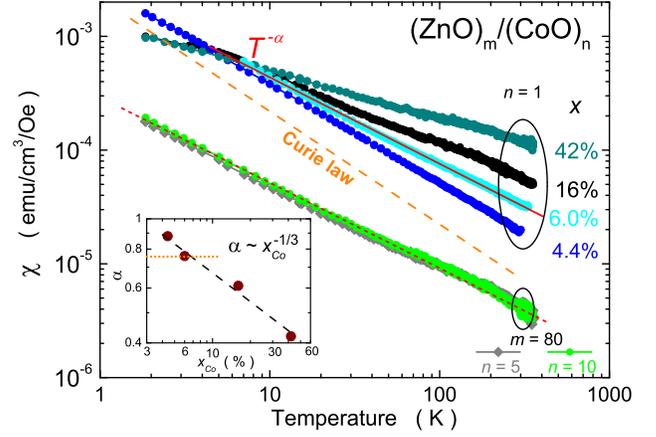}
      \caption{(Color online)   Temperature dependence of magnetic susceptibility. The data sets are labeled by the Co concentration $x$, established for digital alloy layers according to a method described in Sec.~\ref{sec:xCoDetermination} and by their periods for two superlattices. The dashed line indicates the Curie law $\chi(T) \propto 1/T$. Solid and doted lines indicate that $\chi(T) \propto T^{-\alpha}$, where $\alpha < 1$, the dependence specific for a random antiferromagnet. The values of $\alpha$ established in this way for the upper bunch of curves are plotted vs. $x$ in the inset. The dashed black line shows $\alpha \propto x^{-1/3}$, and serves as a guide for the eye. This trend is not obeyed the samples grown in a superlattice fashions, for which the magnitude of $\chi$ is low ($x \approx 0.7$\%), whereas  $\alpha$ (marked as dotted line in the inset) corresponds to $x \approx 6$\%.}
     \label{fig:ChiLogLog}
\end{figure}

We start our quantitative analysis from magnetic susceptibility $\chi(T)$, whose temperature dependence  is obtained  from $M(H,T)$ data measured below 10~kOe.
The values of $\chi(T)$ determined in this way are plotted in a doubly logarithmic scale  in Fig.~\ref{fig:ChiLogLog}.
The data indicate that  $\chi(T) \propto T^{-\alpha}$,  where $\alpha < 1$ and its magnitude   decreases with $x$, as shown in the inset.
Such a dependence is characteristic for random antiferromagnets, that is for paramagnetic compounds with a wide spectrum of (antiferromagnetic) exchange integrals.\cite{Bhatt:1986_PS,Anderson:1986_PRB}
The upper bound of this spectrum, i.e., an effective nearest-neighbor exchange integral $J_{\text{eff}}$ can be evaluated by analyzing $\chi^{-1}$ vs.~$T$ in terms of the high temperature expansion, Eq.~\ref{eq:CW}.

\begin{figure}[bh]
		\centering
        \includegraphics[width=8.5cm]{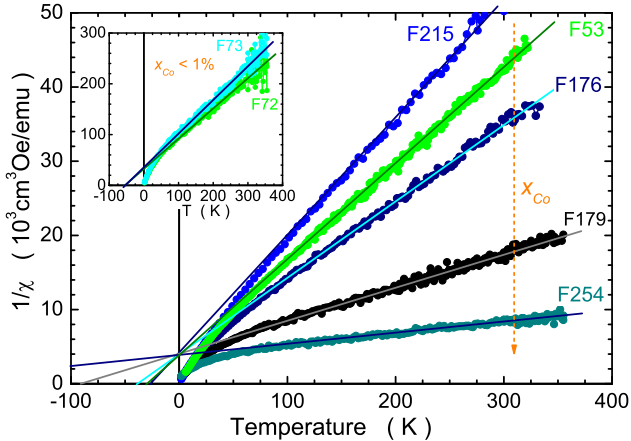}
      \caption{(Color online) The inverse of the magnetic susceptibility $\chi$ as a function of temperature for paramagnetic (Zn,Co)O layers. The main panel groups data for layers in which CoO layers were introduced in a digital way, whereas the inset presents data for the two layers grown in a supperlattice-like fashion. Note the substantially different values of  $\chi^{-1}(T=0)$ for the two types of samples.}
     \label{fig:SQUID_InvChi}
\end{figure}

As shown in Fig.~\ref{fig:SQUID_InvChi}, the inverse of $\chi(T)$ points to a negative sign of the Curie-Weiss temperature $\Theta$, which reconfirms the  antiferromagnetic character of spin-spin interactions.
Slopes of $1/\chi(T)$ vs. $T$ dependencies provide the values of Co concentrations $x$ listed in Table~\ref{tab:samples}. Furthermore,  according to Eqs.~\ref{eq:CW} and \ref{eq:Theta}, the extrapolated values $\chi^{-1}(T=0)$ for randomly distributed spins should be independent of $x$ and directly proportional to the exchange integrals characterizing spin-spin coupling.
And the data gathered in the main part of Fig.~\ref{fig:SQUID_InvChi} instruct us that this is the case for the layers grown in the digital alloy fashion, indicating that independently of the Co content such a growth mode leads to films with randomly distributed Co cations over the ZnO host lattice.


\subsubsection{Determination of Co concentrations}
\label{sec:xCoDetermination}

Turning now to magnetic assessment of an average Co concentration we want to start from a comment concerning a general reliability of the values of $C$ and $\Theta$ determined from the high-$T$ susceptibility.
Our numerous experimental and numerical tests showed that whereas the slope of $\chi^{-1}(T)$ changes substantially in response to small variations of the compensating (substrate) data that the credibility of $\Theta$ value established by this method is much greater than that of $C$.
So, in order to establish more trustworthy values of $x$ in our layers we refer back to the room temperature high-field $M(H)$ data which are far less sensitive to various experimental artifacts and procedure inaccuracies than the determination of the high temperature susceptibility.
Therefore we take $\Theta_0$ from Fig.~\ref{fig:SQUID_InvChi} and substitute the high-field slope of the $M(H)$ in Fig.~\ref{fig:SQUID_2} for $\chi$ in  Eq.~\ref{eq:CW}.
Obtained this way values of $x$ are listed in Table~\ref{tab:samples} as our prime magnetic estimates of Co concentration in the paramagnetic layers.
Noteworthy, we find both sets of 'magnetic' $x$ fairly equivalent each other (most of the discrepancies are well contained within 10\% experimental error) and that both compare favorably to other  methods' estimates (SIMS, EDX, and EPMA).
The only noticeable discrepancy is seen for F254, the most Co-concentrated layer, for which $\Theta \simeq -200$~K and so the high temperature expansion (Eq.~\ref{eq:CW}) ceases to be valid, as the representation of the data \emph{via }the Curie-Weiss law is valid only when $|\Theta |$ is substantially smaller than the temperature of the measurement.

We finally conclude this part stating that the reported here agreement boosts our confidence in the overall correctness of the applied here SQUID-data analysis, in particular that other rendered such an experimental procedure impossible.\cite{Sati:2007_PRL} Furthermore, the data points to the random distribution of magnetic ions, which indicates that interdiffusion is efficient enough to homogenize the Co distribution digitally doped layers.

\subsubsection{Long period (ZnO)$_m$(CoO)$_n$ superlattices}
\label{sec:SuperlatticeLayers}

It appears to be an altogether different story with the long period $m=80$, $n=5$ and 10 superlattices. As evidenced in Figs.~\ref{fig:ChiLogLog} and \ref{fig:SQUID_InvChi}
their $\chi(T)$ stick out completely from those of the layers grown in the digital alloy manner. In fact, a simple simulation for F72 sample shows that the absolute values of $\chi(T)$ can be well reproduced assuming 8:1 partition ratio between pure ZnO and  Zn$_{1-x}$Co$_x$O sublayers with $x$ = 6\% in a rectangular ZnO/Zn$_{1-x}$Co$_x$O superlattice, indicating that interdifussion is not strong enough to homogenize the Co content along the growth direction in such long period superlattices.

\subsubsection{Magnetic anisotropy}
\label{sec:Mgn_Anisotropy}

We examine samples for which XRD measurements (Fig.~\ref{fig:XRD_Anisotropy} in Sec.~\ref{sec:samples}) have revealed the wealth of various $c$-axis arrangements. These are textured films with the dominant orientation of the $c$-axis either along the growth direction (F53) or perpendicular to it (F72) as well as polycrystalline samples  (with randomized $c$-axis orientations) of a medium (F175) or a high Co content (F179), $x = 6.4$ and 16\%, respectively.

\begin{figure*}[th]
		\centering
        \includegraphics[width=17 cm]{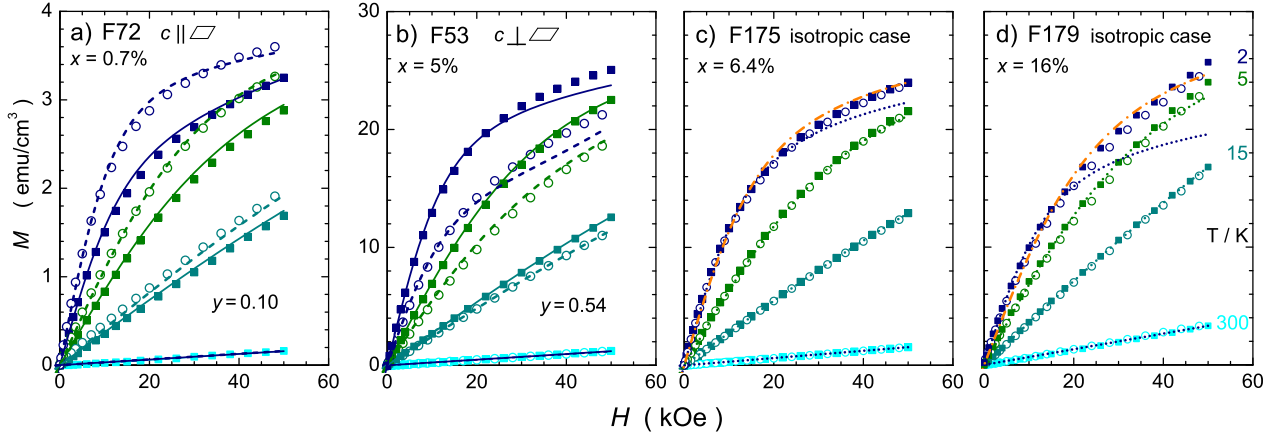}
      \caption{(Color online) Magnetization for a series of samples measured at 2, 5, 15 and 300~K (from top to bottom, indicated in panel (d)) for two orientations of the magnetic field: in-plane (solid squares) and perpendicular to the sample plane (open circles). Lines show modeling of the data by theory outlined in Sec.~\ref{sec:SQUD_theory}. The solid and dashed lines in panels (a,b) are calculated for the mixed anisotropy case considering the in-plane and perpendicular magnetic field, respectively.  The fraction $y$ of grains with the $c$ axis perpendicular the film plane and the effective Co concentration $x_{\text{eff}}(T)$ are the fitting parameters, whose values are displayed in the panels (a,b) and in Table~\ref{tab:xEff}, respectively.  The dotted lines in panels (c,d) represent the isotropic (randomized with respect to $c$-axis) values of magnetization in polycrystalline films. The orange dashed-doted lines represent a considerably improved fit with an effective temperature $T_{\text{AF}}$ added to the calculations performed at the lowest temperature and displayed in Table~\ref{tab:xEff}.}
     \label{fig:SQUID_MatLowT}
\end{figure*}

The experimental data shown in Fig.~\ref{fig:SQUID_MatLowT} (symbols) for parallel and perpendicular directions of the magnetic field  with respect to the film plane demonstrate that magnetization anisotropy  follows the trend expected theoretically (lines) for the distribution of c-axis revealed by the XRD measurements. In particular, in the textured films the magnitude of magnetization is larger for the magnetic field perpendicular to the prevailing direction of the $c$-axis [Fig.~\ref{fig:SQUID_MatLowT}(a,b)]. The description of these data has been carried out following the model presented in Sec.~\ref{sec:SQUD_theory}, where the parameter $y$  quantifying the fraction of the crystal grains having their $c$-axis along the growth direction has been introduced.
According to the fitting procedure, it attains the value of 0.54 in sample F53, for which XRD reveals that the $c$ axis is mostly out of the film plane, whereas $y = 0.10$ in the case of sample F72, where the $c$ axis is preferably oriented in-plane of the film.
In contrast,  magnetization is isotropic [Fig.~\ref{fig:SQUID_MatLowT}(c,d)] in the case of samples that are polycrystalline according to the XRD results.

Finally, we want to underline a value of the presented here analysis of the low temperature $M(H)$ as an accurate method for determining the degree of the crystallographic order in polycrystals exhibiting a strong magnetocrystalline anisotropy.
One has to however keep it in mind that experimental handling of very thin layers is tricky at any circumstances,
and despite a quite strong magnetic anisotropy in (Zn,Co)O, a "slight" error in the substrate contribution can sizable change the input data for fitting and so invalidate the magnetic assessment of the $c$-axis ordering.

\subsubsection{Quantifying spin-spin interactions}

The high temperature data on magnetic susceptibility $\chi(T)$
discussed previously (Fig.~\ref{fig:SQUID_InvChi}) demonstrates that in the films in question Co ions are distributed randomly. Furthermore,  extrapolation  of $\chi^{-1}(T)$  to zero provides magnitudes of Curie-Weiss temperatures for particular samples. Now, assuming that the Co-Co coupling can be characterized by one exchange energy describing the interaction with twelve nearest neighbors, for the experimental value of $\chi^{-1}(T=0) = -\Theta_0/C_0 = 4200 \pm 400$~cm$^{\text{3}}$/emu we obtain  $\Theta_0=-700 \pm 70$~K and $J_{\text{eff}}= 47\pm5$~K.
This value compares very favorably with 42, 51, and 53~K obtained for single crystalline thin films\cite{Sati:2007_PRL} (when we restore the omitted factor "2" in their definition of $J_{\text{nn}}$), single phase bulk crystalls\cite{Kolesnik:2004_JAP}  and powders\cite{Yoon:2003_JAP} (after a necessary adjustment of the material parameters to the values adopted in this study), respectively.

Interestingly, all these values are nearly 40\% smaller than that established for bulk wz-(Cd,Co)S (Ref.~\onlinecite{Lewicki:1990_PRB})
and twice smaller than those found in bulk zinc-blende (Zn,Co)S and (Zn,Co)Se (Ref.~\onlinecite{Lewicki:1989_PRB}). This implies another dependence of $|\Theta_0|$ on the bond length $d$ comparing to the case of Mn-based II-VI DMS, where $\Theta_0$ tends to decrease monotonically  with $d$.\cite{Kolesnik:2004_JAP,Spalek:1986_PRB,Szuszkiewicz:2006_PRB}

The quantitative description of $\bm{M}(\bm{H},T)$ provides also the values of $x_{\text{eff}}$ and $T_{\text{AF}}$, collected in Table~\ref{tab:xEff}.  These parameters supply also information on spin-spin coupling. In particular, the magnitudes of $x_{\text{eff}}$ determined at 2~K are in agreement with the concentrations of Co ions having no other Co as the nearest neighbor (see, Sec.~\ref{sec:SQUD_theory}).
This means that there is a strong antiferromagnetic coupling between Co ions occupying any of the twelve nearest neighbor positions in the cation sublattice. At the same time, much lower values of  $T_{\text{AF}}(2$~K), comparing to high temperature Curie-Weiss $\Theta$, indicate that coupling to next nearest neighbors is relatively weak, which points to a rather short range character of the antiferromagnetic spin-spin interactions.

Altogether, the findings accumulated so far support those { \em ab initio} simulations which do {\em not} predict the presence of ferromagnetic interactions at any distance of Co-Co pairs.\cite{Gopal:2006_PRB,Lany:2008_PRB,Iusan:2009_PRB}

\begin{table}
\begin{tabular}{c c c c c}
\hline\hline
$T$(K) & F72 & F53 & F175 & F179 \\
 & \hspace{0.4cm}--\hspace{0.4cm} & \hspace{0.4cm}5.0\%\hspace{0.4cm} & \hspace{0.4cm}6.4\%\hspace{0.4cm} & \hspace{0.4cm}16\%\hspace{0.4cm} \\
\hline\hline
300& 1.00 & 1.00 & 1.00 & 1.00 \\
15& 0.51 & 0.61 & 0.45 & 0.27 \\
5& 0.50 & 0.53 & 0.38 &	0.19 \\
2& 0.49 & 0.51 & 0.35 &	0.14 \\
$2+T_{\text{AF}}$& -- & -- & 0.37 &	0.18 \\
\hline
singles & 0.92 & 0.54 & 0.45 & 0.14\\
\hline
$T_{\text{AF}}(2$~K)& -- & -- & 0.35~K & 1.45~K \\
\hline\hline
\end{tabular}
\caption{Temperature dependence of the effective Co concentration $x_{\text{eff}}$, given here as a ratio to its value $x$ at 300~K, for the four samples presented in Fig.~\ref{fig:SQUID_MatLowT}.  This is followed by a line giving the statistically expected fraction of single Co cations present in the wurtzite lattice at given $x$, corrected in the case of the superlattice sample (F72) by the period ratio (Sec.~\ref{sec:SuperlatticeLayers}).
The bottom line lists the values of effective temperatures $T_{\text{AF}}$ added to the calculations of $\bm{M}(\bm{H},T)$ for two samples with the highest $x$ (F175 and F179) at 2~K. }
\label{tab:xEff}
\end{table}

\subsubsection{Spin-glass freezing}
\label{sec:Spin-glass}

\begin{figure}[t]
		\centering
        \includegraphics[width=8.5cm]{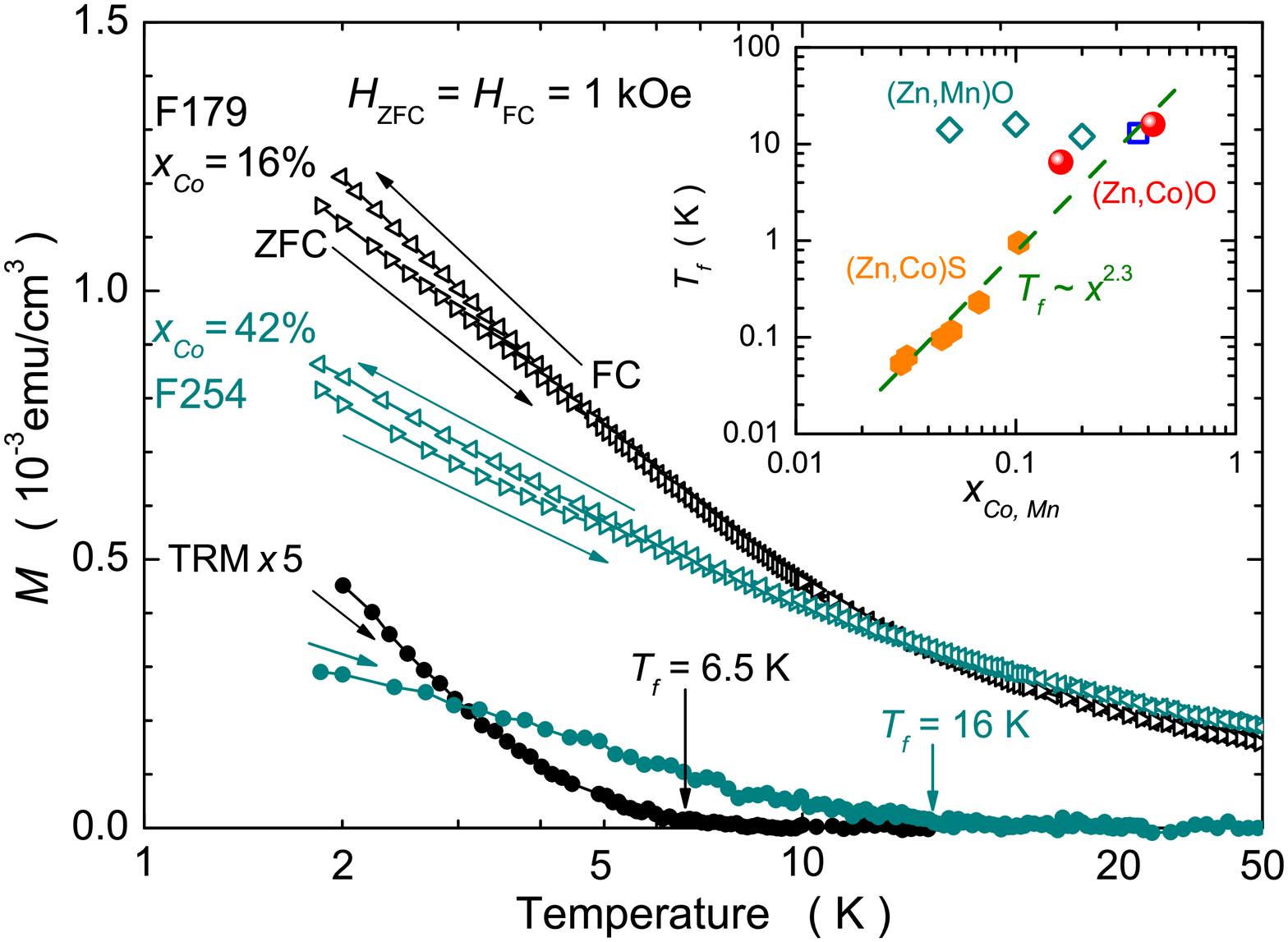}
      \caption{(Color online) Main figure: a spin-glass-like behavior of two paramagnetic layers with the highest Co content. Darker color: F179, $x \cong 16$~\%; lighter one: F254, $x \cong 42$~\%. Zero field cooled (ZFC) and field cooled (FC) measurements performed at 1~kOe are indicated by arrows and marked by open triangles pointing towards increasing or decreasing temperature, respectively. Full circles mark thermoremanent magnetization, TRM (magnified 5 times for better presentation), which enables us to establish the spin-glass freezing temperature, $T_f$, most reliably. Inset shows $T_f$ dependence on the transition metal content: $x_{Mn}$ in (Zn,Mn)O (open diamonds from Ref.~\onlinecite{Kolesnik:2002_JS:INM} and open square from Ref.~\onlinecite{Fukumura:2001_APL}) and $x$ in (Zn,Co)S (full hexagons from Ref.~\onlinecite{Shand:1991_PRB} and red bullets in (Zn,Co)O (this study).}
     \label{fig:SG_M-T}
\end{figure}

Since the very early stage of magnetic studies of DMSs, a low temperature cusps on $\chi(T)$ curves were observed over a broad range of magnetic cation concentration.\cite{Galazka:1980_PRB,Mycielski:1984_SSC,Twardowski:1987_PRB} Despite the lack of competing interactions (at least in the compounds where an effective p-type doping is far too low to initiate a carrier mediated ferromagnetic ion-ion coupling) but on the account of the presence of the other two key ingredients, namely positional disorder and spin frustration, these materials exhibit spin-glass characteristics driven entirely by antiferromagnetic interactions.\cite{Mydosh:1993,Jaroszynski:1998_PRL}
And our two paramagnetic layers with the highest Co content show a weak low temperature feature that closely resembles previous findings in other DMS compounds.\cite{Galazka:1980_PRB,Mycielski:1984_SSC,Twardowski:1987_PRB}
Figure~\ref{fig:SG_M-T} collects zero-field cooled (ZFC), field-cooled (FC) and thermoremanence (TRM) magnetization for these layers.
A typical for a transition to a glassy state bifurcation on ZFC and FC curves is seen.
However a strong paramagnetic background masks the effect of freezing (characteristic cusp on the ZFC) considerably, therefore the TRM measurement serves to accurately establish the freezing temperature $T_{\text{f}}$ for these layers.

An important question arises now whether the presented here history dependent effects and a very weak magnetic hysteresis developing at $T < T_{\text{f}}$ (not shown) are indeed related to the spin-glass-like freezing, or they are a manifestation of the dynamical blocking of some small ferromagnetic clusters (Co-metal droplets or some Co-rich spinel precipitates) that together give rise to superparamagnetic-like behavior.
These doubts are fueled by our earlier observation that a contribution from metallic Co is present in the XPS spectra both of the PM layer F179, one of these large $x$ layers which show the freezing.
In order to resolve this issue one should resort to studies of the long time scale dynamics of the non-reversible part of the signal.\cite{Mydosh:1993}
However, the presence of a large PM background renders such an experiment problematic.

In an attempt, we have found a lack of any noticeable time dependence of ZFC magnetization up to $2 \cdot 10^4$~s after the $x = 16$\% sample was zero-field-cooled to 5~K, that is below its $T_{\text{f}}$ (not shown),
what is in strike difference with fast relaxations observed in ZFC state for superparamagnetic (SP) layer F309, presented in Sec.~\ref{sec:BulkSP}, however on the account of a very small accuracy, we regard this finding as rather weak evidence.

Therefore, we are going to argue for the spin-glass freezing on the account of the minute magnitude of $M$, corresponding solely to the PM response of these two samples.
We note, that if sizable and numerous Co-rich metallic (ferromagnetic) inclusions were present they would dominate the magnetic response at weak magnetic fields (at any temperature) considerably increasing the magnitude of $M$, see ~{\em e.g.},~Ref.~\onlinecite{Karczewski:2003_JSNM}).
Indeed, as it is exemplified in Sec.~\ref{sec:BulkSP}, the FC magnetization of this aforementioned SP layer F309 is found be nearly 500 times larger at low temperatures than in the presented here layers, despite ten times weaker magnetic field used during the measurement of the SP sample.
Therefore, predominantly on the account on the size of the irreversible effects, we rule out their superparamagnetic origin and adopt the spin-glass explanation.

Finally, we find a good correspondence between the observed here freezing temperatures with those reported for similar systems.
As indicated in the inset to Fig.~\ref{fig:SG_M-T} found here $T_{\text{f}}$ follow the already established exponential dependency $T_f \sim x^{\alpha}$ with $\alpha = 2.1 \pm 0.4$ for bulk (II,Co)VI compounds\cite{Shand:1991_PRB} and $\alpha \simeq 2$ for the other legacy DMSs.\cite{Twardowski:1986,Twardowski:1987_PRB}
Interestingly, $T_{\text{f}}$ data from Ref.~\onlinecite{Kolesnik:2002_JS:INM} do not joint none of these trends indicating that an unidentified factor must have contributed to the observed there freezing-like behavior.

\subsection{Bulk superparamagnetic contribution}
\label{sec:BulkSP}

We find that the total amplitude of $M(T,H)$ can, in general,  be decomposed into three main components: (i) the already discussed paramagnetic (PM) term brought about by randomly distributed Zn-substitutional Co cations;  (ii) a relatively slowly saturating in the magnetic field \emph{superparamagnetic} (SP) component, and (iii) a fast saturating, highly anisotropic and temperature independent up to 300~K \emph{ferromagnetic} (FM) contribution.

The SP component is present in all our samples grown at 200$^\circ$C and above. Additionally, in such samples a FM contribution may show up. We postulate now, and elaborate it later, that it is few nm thick mesh of metallic cobalt present on the layer/substrate interface which is responsible for the ferromagnetic component.
The existence of such a film has already been documented in the previous sections devoted to general characterization of our layers (SIMS, TEM, XPS, and XMCD).
We also regard this interface FM to be completely temperature independent, as  finite-size effects are not supposed to show up for Co film thickness above $\sim 1.5$~nm.\cite{Ney:2000_PRB}

Conversely, the PM and SP components are assumed to stem from the bulk of the layer.
So, we allow all the Co atoms which build up the volume of the layer to either take random cation sites, giving rise to the PM response, or to aggregate into nano-crystallites of a large magnetic 'supermoment' $m$, responsible for the SP part of $M(T,H)$.

\begin{figure}[b]
		\centering
        \includegraphics[width=8.5cm]{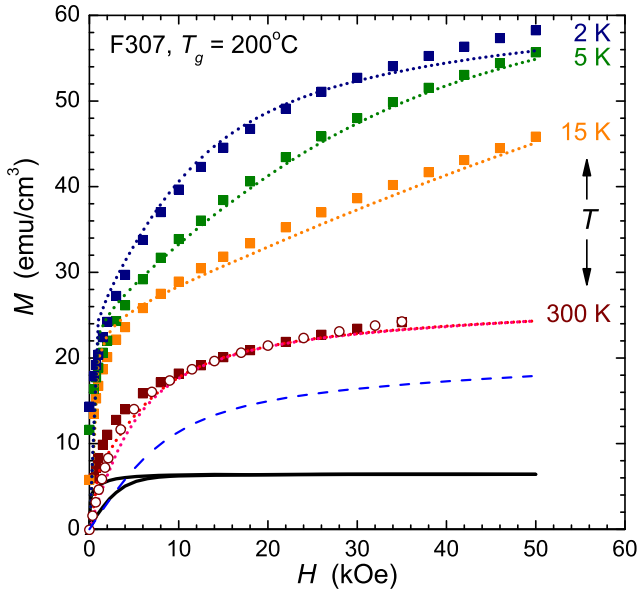}
      \caption{(Color online)  A representative (volume) magnetization for a sample grown at 200$^\circ$C (F307). Solid squares are obtained  with the magnetic field applied in the sample plain. For clarity the data obtained in perpendicular configuration are shown only for 300~K (open circles). Dotted lines indicate results of the modeling which assumes a presence of three independent magnetic contributions: a paramagnetic, a superparamagnetic, and a ferromagnetic one, as detailed in the text. The dashed line represents the magnitude and room temperature curvature of the superparamagnetic component, whereas the two solid lines represent a derived from sample F328 the anisotropic ferromagnetic component used in the modeling (see Sec.~\ref{sec:FM-samples}).}
     \label{fig:M307Exmaple}
\end{figure}

Dotted lines in  Fig.~\ref{fig:M307Exmaple} indicate a result of the modeling of the experimental data by a straight sum of these three magnetic components.
We calculate the PM response following the method described in the previous section.
In particular, we use isotropic paramagnetic $M(T,H)$, allow for the temperature dependence of $x_{\text{eff}}$,
and take $T_{\text{AF}} =0$ for simplicity.
The magnitude of this signal is parameterized by the concentration of randomly distributed Co cations, $x_{\text{eff}}(300$~K).

As described in the next section, the FM component has been derived from the room temperature $M(H)$ of the layer F328 and scaled to this sample according to their areas, so there is no free parameters here.
The fitting itself is performed only for $H > 5$~kOe, as we do not have enough knowledge on the processes which determine coercivity and remanence and hence the magnitude of $M$  at weak magnetic fields.

{\em A priori}, as mentioned in the Introduction,  spins contributing to the SP signal might originate from two distinct kinds of systems: (i) nanoparticles of metallic Co, CoZn, or related intermetallic ferromagnetic compounds; (ii) uncompensated spins residing at the surface of {\em antiferromagnetic} wz-CoO nanocrystals, as observed in the case of nanoparticles of NiO.\cite{Proenca:2011_PCCP} In the latter case, involving magnetically inert Co atoms in  wz-CoO nanocrystals, the total Co density should be larger than the one seen magnetically.

The SP contribution to $M(T,H)$ is approximated by the Langevin function $mN_{\text{SP}}L(x)$, where  $x = m H / k_{\text{B}} T$ and $m =n\mu_{\text{Co}}$. Guided by XPS and TEM data, as a first and successful guess we assume that nanoparticles constitute of Co metal. In such a case,  $\mu_{\text{Co}} = 1.7$~$\mu_{\text{B}}$ and the number $n$ of Co atoms giving rise to the magnetic moment $m$  and the density of nanoparticles $N_{\text{SP}}$ are bound together by the requirement that the total density of the Co atoms, $N_0x_{\text{eff}}(300$~K$) + nN_{\text{SP}}$ is equal to the Co concentration found by SIMS in the volume of this layer, $x^{\text{SIMS}}(\text{F307}) = 8$\%.
Therefore, out of three quantities, $n$, $N_{SP}$, and $x_{\text{eff}}(300$~K), we are left with two temperature and field independent adjustable parameters.

\begin{figure}[t]
		\centering
        \includegraphics[width=8.5cm]{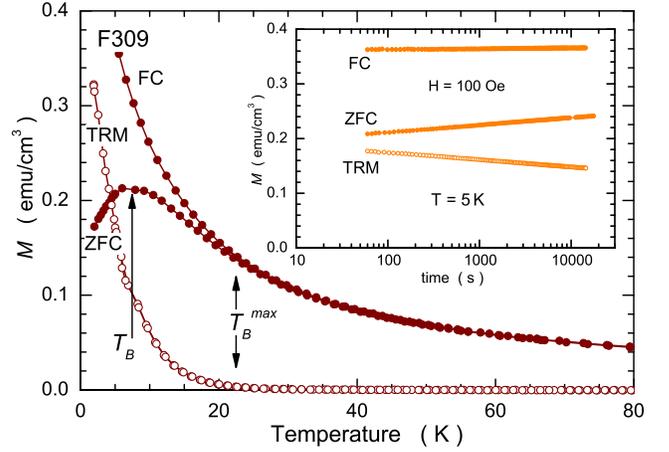}
      \caption{(Color online) Blocking phenomenon in a superparamagnetic sample (F309). Main part: temperature dependence of the zero field cooled (ZFC), field cooled (FC) and thermoremanent (TRM) magnetization. Arrows indicate positions of the blocking temperature ($T_{\text{B}}$) and the maximum blocking temperature ($T_{\text{B}}^{\text{max}}$). Insert: time evolution of the magnetization in these three magnetic states measured at 5~K (below $T_{\text{B}}$).  }
     \label{fig:F309SP}
\end{figure}
The fitting results presented in Fig.~\ref{fig:M307Exmaple} demonstrates that 65\% of Co present in the bulk of the layer assumes ZnO lattice cation sites indicating that this is still the preferred site for Co atoms.
The remaining 35\% of Co precipitates into nanoparticles containing (on average) $n \simeq 750$ Co atoms (approx.~2.5~nm across) if they indeed assume the form of bulk Co.
The magnitude of this superparamagnetic component at 300~K is presented in Fig.~\ref{fig:M307Exmaple} as a dashed line.
As seen, this is the superparamagnetic contribution that is responsible for the pronounced high temperature curvature and slope, whereas the paramagnetic component governs the low temperature magnitude of $M$ at high magnetic fields.
The interface-related ferromagnetic component accounts for the quickly magnetizing and anisotropic component.
Finally, we note that obtained this way such a good reproduction of the experimental data  strongly indicates that all these three components are to the first order independent one from another.

For further studies of the SP contribution, particularly to examine blocking properties, we select a sample with no FM component. The relevant findings are presented in  Fig.~\ref{fig:F309SP}, where a clear and quite sharp maximum of ZFC magnetization  and a bifurcation between ZFC and FC magnetizations are seen at about $T_{\text{B}} =7$ and $T_{\text{B}}^{\text{max}}=22$~K, respectively.
These blocking temperatures allow to assess an average and maximum size of statistically relevant nanocrystals.
If surface anisotropy\cite{Jamet:2001_PRL} is unimportant, we can take  magnetocrystalline anisotropy density $K$ of bulk cobalt, $K=5 \times 10^6$ erg/cm$^3$. Employing the widely accepted formulae for DC type SQUID magnetometry, $T_{\text{B}} \simeq K V / 25 k_{\text{B}}$, where $V$ is the volume of the nanoparticle, we obtain the average and the statistically relevant maximum size of Co nanocrystals as 2.4 and 3.3~nm, respectively.
Both values correspond precisely to our previous estimate of 2.5~nm, obtained from Co counting in the different layer.

Finally, we confirm the SP nature of the magnetic  signal in question by examining relaxation of ZFC, FC, and TR magnetizations below $T_{\text{B}}$, as shown in the inset to Fig.~\ref{fig:F309SP}.
The lack of noticeable relaxation of the FC signal and relatively fast relaxations of both ZFC and TR magnetizations, upwards towards the FC value and downwards towards zero, respectively, agree with the general expectations for the time evolution of these three type of magnetization in a blocked state of superparamagnets.

\subsection{Interfacial ferromagnetism}
\label{sec:FM-samples}
\

\begin{figure}[b]
		\centering
        \includegraphics[width=8.5 cm]{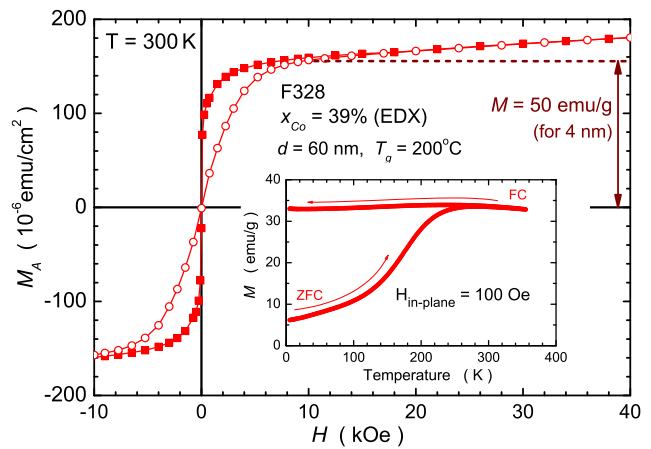}
      \caption{(Color online) Magnetization of layer F328. Main part: magnetic moment per unit area, $M_{\text{A}}$, at 300~K. Solid squares are obtained  with the magnetic field applied in the sample plain, open circles mark the data obtained in perpendicular configuration. The dashed horizontal line indicates the estimated saturation level of the ferromagnetic part of the data which we use to construct a prototype data sets for this magnetic contribution. Inset presents low field in-plane zero field cooled (ZFC) and field cooled (FC) magnetization. }
     \label{fig:MareaF328}
\end{figure}
On the account of the already presented  XPS, HR-TEM and XMCD studies we unambiguously assign the swiftly magnetizing and anisotropic signal to the evidenced there few nm thin layer of metallic Co mesh and we will show that indeed it shows many characteristics corresponding to bulk cobalt.

In Fig.~\ref{fig:MareaF328} we plot a 'sheet' magnetization  $M_{\text{A}}$, the magnetic moment per unit area, obtained at 300~K for layer F328, the thinnest layer of all studied layers and the main subject of the aforementioned characterization efforts.
The dashed line marks in the figure the expected saturation level, and these are actually the points taken from this line plus the original data taken from $-10 < H < 10$~kOe field range (for both orientations of $H$) which are used in the modeling of the $M(T,H)$ presented as the solid lines in Fig.~\ref{fig:M307Exmaple}.
Interestingly, similarly to layer F307, we can  completely remove this ferromagnetic component from experimental data in some other layers (in both configurations of $H$) by a simple subtraction of these  established here prototype $M(H)$, providing we scale the data according to the layers' areas first.

However, the TEM cross-section images of the F328 interface area (see Fig.~\ref{fig:TEM}b and Fig.~\ref{fig:ChemAnalysis}) indicate some kind of a granular form of this interfacial Co film.
Moreover, our XMCD studies (see Sec\ref{sec:XMCD}) of the interfacial part of the layer also point to a granual form of the Co layer at the interface.
So, the important question arises whether these Co granules are (magnetically) independent or whether they coalesce towards a uniform Co layer.
In order to distinguish between these two possibilities we perform zero field cooled (ZFC) and field cooled (FC) measurements at weak magnetic field.
A presence of a well developed maximum on the ZFC curve associated with a lack of a corresponding one on the FC magnetization would indicate  the temperature dependent (superparamagnetic) blocking phenomenon (as already evidenced for a superparamagnetic layer F309 in Fig.~\ref{fig:F309SP}),
and so it would indicate a loose distribution of cobalt granules on the interface.
But, as indicated in the inset to Fig.~\ref{fig:MareaF328}, we find no real maximum what rules out the individual (super-)moments scenario in favor for the coalescing picture, and so pointing strongly to a percolating  mesh formed from closely packed and interconnected Co granules.

Another support for the Co mesh comes from the notion that when we recalculate $M_{\text{A}}$ onto $M$ (volume), we obtain $M \simeq 500$~emu/cm$^3$ for a reasonable value of the interface film thickness, $d_{\text{Co}}=4$~nm.
This is only a third of a bulk cobalt saturation magnetization what indicates that indeed Co granules occupy only a (small) fraction of the interfacial region.
We also note that the established granular form of the Co interfacial film does not contradict our initial assumption about the lack of finite-size effects.
In this point we share the view of Ney and coworkers\cite{Ney:2010_NJP}  that Co granules which are 3-5~nm big should not show any pronounced reduction of the magnetization between helium and room temperature.

Finally, we note that it is the flatness of this Co film (mesh) that is responsible for the universality of the easy-plane magnetic anisotropy.
However, the experimentally observed anisotropy field, $H_{\text{A}}^{\text{exp}} \simeq 5$~kOe is about 3 to 4 times weaker than the expected for a perfectly flat and uniform Co film, $H_A^{sh} \simeq 18.5$~kOe.
We take this anisotropy weakening as yet another manifestation of the patchy character of the Co film, and treat the correspondence of the magnitude of $H_{\text{A}}^{\text{exp}}$ to the magnitude of Co  magnetocrystalline anisotropy field as purely coincidental, predominately due to the lack of crystallographic ordering of the Co granules seen in TEM images.


\section{Microwave conductivity}
\label{sec:MW}

The presence of metal Co inclusions visualized by
XPS, as described above, while invisible in standard conductivity
measurements, could lead to detectable dielectric losses. In
particular, the microwave cavity perturbation
technique,\cite{Champlin:1961_IRE,Brodwin:1965_JAP} that is the
observation of modifications of the cavity resonance by the
presence of a sample, may serve to evaluate the magnitude of
microwave conductivity $\sigma_{\text{AC}}$ of our films and to find out if it
correlates with their magnetism .

\subsection{Theory of microwave cavity perturbation technique}
We use microwave cavity perturbation technique, first developed to
measure the complex dielectric permittivity of small spherical
samples placed in a microwave
cavity.\cite{Champlin:1961_IRE,Brodwin:1965_JAP} This method was
then adopted to the case of ellipsoidal
samples,\cite{Buravov:1971_IET} and subsequently generalized to
cover a wide range of conductivity magnitudes, from the
quasi-static case up to the skin-depth
regime.\cite{Jaworski:1977_B,Kleiner:1993_IJIMW} In most cases the
cavity perturbation method has been applied to uniform samples
only. One of a few exceptions is the consideration of a layered
structure consisting of a thin conducting film deposited on a
dielectric substrate.\cite{Gregorkiewicz:1980_RE}

In order to apply this method to thin films of (Zn,Co)O, we have
developed a more rigorous and, simultaneously, more general
theoretical approach.
\begin{figure}[h]
   \centering
    \includegraphics[width=8cm]{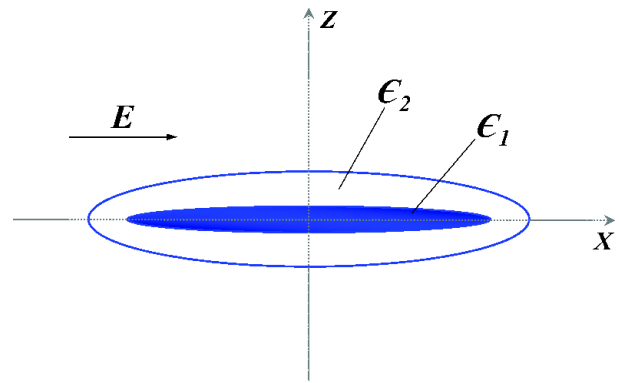}
    \caption{(Color online) Spheroidal model of a two-layer sample. The electric field
    component of the microwave electromagnetic radiation is parallel to the
    longest axis of the sample. The $\epsilon_1$ is the film dielectric permittivity,
    $\epsilon_2$  is the substrate dielectric permittivity. }
    \label{fig:MW_1}
\end{figure}
We consider spheroidal samples consisting of two layers (the film
and the substrate), as shown schematically in Fig.~\ref{fig:MW_1}.
We adopt the quasi-static
approximation\cite{Gregorkiewicz:1980_RE}, {\em i.~e.} we assume
the effective wavelength of the microwave radiation within the
layer to be much greater than its transversal dimension. Moreover,
we assume that a relative deviation from the cavity resonance
frequency $\omega_0$ is small, however the perturbed microwave
electric fields $E_1$ and $E_2$ within particular layers may
differ considerably from the value $E_0$ for the empty cavity. The
fundamental perturbation formula, describing the modification of
cavity resonance by the presence of the sample, can be then
concisely written as,
\begin{equation}
\delta -\mbox{i}\Delta/2 = \frac{\alpha_1[\epsilon_1(\omega)-1] +
\alpha_2[\epsilon_2(\omega)-1]}{1 + n_1[\epsilon_1(\omega)-1] +
n_2[\epsilon_2(\omega)-1]}, \label{eq:MW}
\end{equation}
where $\delta = (\omega_0 - \omega)/\omega$ and $\Delta = 1/Q -
1/Q_0$ denote a relative shift from the values of $\omega_0$ and
the quality factor $Q_0$ brought about by the sample; the indices
1 and 2 refer to the film and the substrate, respectively;
$\alpha_i = V_i|E_{0\text{s}}|^2/(2\int \mbox{d}v|E_0|^2)$, where
$V_i$ are the sample-related volumes while the integration extends
over the whole cavity; $E_{0\text{s}}$ is the value of $E_{0}$ at
the sample position; $\epsilon_i(\omega) =\epsilon'_i(\omega)
-\text{i} \epsilon''_i(\omega)$ are complex dielectric
permittivities; and $n_i$ are depolarization
factors.\cite{Osborn:1945_PR} It can be easily seen that
Eq.~(\ref{eq:MW}) is a direct generalization of a similar formula
derived earlier for a single layer.\cite{Buravov:1971_IET}

Assuming that the substrate parameters are known, one can
determine the film permittivity $\epsilon_1(\omega)$ from
experimental values of $\delta$ and $\Delta$. For a strongly
conducting film the expression describing  $\epsilon'_1(\omega)$
is numerically unstable, {\em i.~e.}, we cannot determine
$\epsilon'_1(\omega)$ reliably if $\epsilon''_1(\omega) \gg
\epsilon'_1(\omega)$. However, Eq.~(\ref{eq:MW}) can be
effectively solved for $\epsilon''_1(\omega)$, leading to the
expression which is analogous to that
derived\cite{Buravov:1971_IET} for a single layer sample and
connects directly the cavity bandwidth with the layer
conductivity. For flat samples and highly resistive substrates the
influence of substrate properties is small and can be neglected to
a good approximation.

The above exact analytical solution for layered media has been
derived for spheroidal samples. In practice, however, one deals
often with samples in the form of discs or rectangular prisms and
the depolarization factors for sample shapes other than
ellipsoidal are known only approximately. Furthermore, the theory
is developed for symmetric structures (Fig.~\ref{fig:MW_1}) in
which the film is placed in the symmetry plane of the spheroid,
while real films are typically deposited close to the substrate
surface. Therefore, replacing a real sample by an equivalent model
ellipsoid in data analysis could be a source of a systematic
error. Fortunately, both numerical analysis, and test measurements
showed\cite{Gregorkiewicz:1979_PhD} that for the film parameters
distinctly different from those of the substrate that error is
small and measurement results only weakly depend on the film
position.

\subsection{Results of microwave studies}
Our measurements are carried out at room temperature employing a
cylindrical cavity of the diameter $d = 49$~mm and height $h =
35.5$~mm, which operates in the $TE_{112}$ mode of resonant
frequency of $\sim 9.2$~GHz. The sample is placed in the maximum
of the electric field in the plane perpendicular to the cavity
axis at the distance $l=h/4$ from the cavity bottom. The resonance
curve (microwave power reflected from the cavity measured at
frequencies near and at the resonance) is sampled for 400
frequency values for the cavity without and with the studied
sample. The fitting to a standard Lorentzian function allows to
determine both the frequency shift and the bandwidth change. The
filling factors $\alpha_1$ and $\alpha_2$ are calculated as the
ratio of the film and substrate volume to an effective volume of
the cavity, dependent on its dimensions and the distribution of
the electromagnetic field for a given resonant mode. It is assumed
that for the flat samples under consideration depolarization
factors $n_i$ are small but nonzero, making then possible to
measure also highly conducting samples.

Measurements of microwave AC conductivity have been performed for a series of (Zn,Co)O films.
Because this method can sensitively detect even small highly conducting objects we search for
correlation between $\sigma_{\text{AC}}$  and layer's magnetic characteristics, since the latter depend strongly on the presence and type of Co-rich inclusions, as it has been elaborated in the previous section.
Importantly, this is a contactless method, what allows to investigate also films which are highly resistive according to
DC measurements.

\begin{figure}[h]
    \centering
    \includegraphics[width=8 cm]{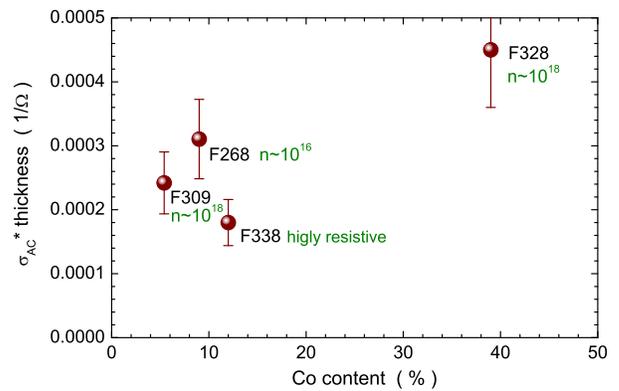}
    \caption{(Color online) Sheet AC conductivity ($\sigma_{\text{AC}}$ times layer thickness) versus EDX Co content. Labels indicate layer's approximate electron concentration (in cm$^{-3}$, see Table~\ref{tab:samples} for more details).}
    \label{fig:Sigma-Co}
\end{figure}

We start the description of the results by noting that no evident correlation between the AC electric conductivity and the DC one is observed for our (Zn,Co)O films.
In particular $\sigma_{\text{AC}}$  does not correlate with the free electron concentration deduced from the Hall effect, as it is indicated in Fig.~\ref{fig:Sigma-Co}.
Neither it correlates with Co content represented in the same figure by its EDX values, as this method  provides us with an averaged Co content in the whole layer, including the Co-enriched interfacial film.
More importantly, and to some extent surprisingly, we see no correlation with the magnetic properties of the layers, however $\sigma_{\text{AC}}$ appears to assume the largest values in samples exhibiting the interfacial ferromagnetism, or to be precise here, in layers having the film of interconnected metallic cobalt granules on the layer/substrate interface (Sec.~\ref{sec:FM-samples}), and so providing an effective conducting medium within the layer.
Three such layers (F268, F328 and F338) are exemplified in Fig.~\ref{fig:Sigma-Co}, when we plot sheet AC conductivity ($\sigma_{\text{AC}}$ times layer thickness) versus the EDX Co content.
We note that expressed this way $\sigma_{\text{AC}}$ changes for these samples only by a factor of 2.5 indicating an important, if not dominant, role of metallic inclusions at the interface in both the determination of the magnitude of $\sigma_{\text{AC}}$ and in the observed discrepancy between DC and AC measurements.
Finally, we note that the observed such a large disagreement between DC and AC conductivity of these layers confirms our previous assessments about a discontinuous character of the interfacial Co layer.

On the other hand layer F309 exhibits a similar magnitude of the sheet AC conductivity despite the fact it shows only the superparamagnetic response (Sec.~\ref{sec:BulkSP}).
In this case the sizeable conductivity stems solely from the large volume of the film (about 20 times larger than of F328) indicating that indeed the method is capable of addressing the tiny and separated metallic (conductive) objects present in the investigated specimen.
However, in this case also, the correct assignment of the experimental finding would not be possible without the joint characterization effort of all the methods employed in this study.

\section{Conclusions}
\label{sec:Concl}
A series of (ZnO)$_m$(CoO)$_n$ digital alloys ($m=2,8,$~$n=1)$ and superlattices ($m=80$,~$n=5,10$) grown by atomic layer deposition have been investigated by a range of experimental methods. The data provide evidences that the Co interdiffusion in the digital alloy structures is sufficiently efficient to produce truly random Zn$_{1-x}$Co$_x$O mixed crystals with $x$ up to 40\%. Conversely, in the superlattice structures the interdiffusion is not strong enough to homogenize the Co content along the growth direction results in the formation of (Zn,Co)O films with spatially modulated Co concentrations. High resolution SQUID measurements have demonstrated that all structures deposited at 160$^\circ$C exhibit magnetic properties specific to dilute magnetic semiconductors with localized spins $S = 3/2$ randomly distributed over cation sites, and mutually coupled by strong, but short range, {\em antiferromagnetic} interactions. The determined magnitude of the exchange energy describing coupling between the nearest neighbor Co pairs corroborates quantitatively its previous determination\cite{Sati:2007_PRL} and qualitatively those first principles studies, in which the spin-spin coupling was found to be antiferromagnetic.\cite{Gopal:2006_PRB,Lany:2008_PRB,Iusan:2009_PRB} The presence of positional disorder within the spin system coupled by antiferromagnetic interactions lead to low temperature spin-glass freezing, not observed earlier in (Zn,Co)O. According to our findings, characteristics of these freezing are similar to those reported for other II-VI DMSs.

Apart from a possible influence of some other growth parameters, it is the growth temperature in excess of 160$^\circ$C that is the decisive factor leading to the occurrence of ferromagnetic-like features in our layers.
Whereas our previous investigations\cite{Godlewski:2011_PSSB} indicated a correlation between accumulation of Co metal at the interface and the observed room temperature ferromagnetic response, in the present work we have investigated the character of this accumulation and verified that indeed the FM response comes from the interface region where Co is accumulated. The SIMS, TEM, and XMCD investigations have indicated that the inclusions are in the form of somewhat oblate (few monolayer thick) Co particles or a flat mesh connected magnetically. The latter is in line with the results of magnetic measurements showing clearly that the ferromagnetic easy axis is in-plane.

The Co mesh at the interface may suggests that the interface acts as a Co sink which stops to operate once the mesh is built. This would explain why the magnitude of the ferromagnetic signal is virtually independent of the thickness of (Zn,Co)O film. Finally, the assignment of the ferromagnetic features to the interfacial layer makes it possible to understand significant deviations from the classical superparamagnetic behavior. Such properties are visible in many high-temperature magnetically doped semiconductors and oxides.\cite{Coey:2010_NJP}
However, the Co layer is not continuous according to transport investigations, as we observed the noticeable difference between the DC and AC (microwave) conductivity that indicates the presence of highly conductive areas within our (Zn,Co)O films but no DC metallic-like conductivity.

In addition to the ferromagnetic signal, particularly well visible in thin samples, in all layers grown at high temperature a superparamagnetic response has been observed. Guided by XPS data, this magnetization component has been assigned to metallic Co-rich nanoparticles disperse over the film volume. This view has been further supported by comparing the magnitude of the total magnetic signal to the Co concentration provided by SIMS.  The important question is whether the presence of Co inclusions results automatically in superparamagnetic-like features. According to the XPS results this is not necessarily the case, as some concentration of metal Co inclusions have been detected in all samples, also in those grown at low temperature and showing only a paramagnetic response. In contrast, in samples grown at higher temperature, presumably because of more efficient aggregation of sizable Co-rich nanoparticles, the superparamagnetic contribution is significant.

In general terms, our results presented here as well as in parallel work on TM-doped nitrides,\cite{Bonanni:2011_PRB,Navarro:2011_PRB} reemphasize the necessity of using a range of nanocharacterization tools in order to assess the lateral {\em and} vertical distributions of TM ions in the studied films. A non-random distribution of TM atoms and a non-random distribution of TM-rich aggregates underline pertinent features of these systems, particularly their magnetic properties. These distributions depend sensitively on growth parameters, which make it possible to obtain a variety of different systems, ranging from DMSs with a random distribution of magnetic cations to novel nanocomposites.

\begin{acknowledgments}
The work was supported by the European Research Council through the FunDMS Advanced Grant (\#227690) within the "Ideas" 7th Framework Programme of the EC and by European Regional Found through blgrants Innovative Economy Operational Programme 2007-2013 (POIG.01.01.02-00-008/08, POIG.02.01-00-14-032/08, and InTechFun: POIG.01.03.01-00-159/08), and  SemiSpinNet (Grant No. PITNGA-2008-215368). EXAFS and XANES research received funding from the European Community's 7th Framework Programme under grant agreement No. 226716.
\end{acknowledgments}

\bibliography{ref_2012_01_13}

\end{document}